\documentclass[a4paper,11pt]{article}
\pdfoutput=1

\usepackage{jcappub} 
\usepackage[T1]{fontenc} 
 
\usepackage{graphicx}
\usepackage{hyperref}
\usepackage{siunitx}
\usepackage{multirow}
\usepackage{txfonts}
\usepackage{subfigure}
\usepackage{float}
\usepackage{bm,array}
\usepackage{amsmath}
\usepackage[dvipsnames]{xcolor}
\definecolor{mygreen1}{rgb}{0.05, 0.55, 0.05}

\title{\boldmath Total power horn--coupled 150~GHz LEKID array for space applications}

\author[a,b,1]{A. Paiella,\note{Corresponding author.}}
\author[a,b]{A. Coppolecchia,}
\author[a,b]{P. de Bernardis,}
\author[a,b]{S. Masi,}
\author[b]{\\A. Cruciani,}
\author[a,b]{L. Lamagna,}
\author[c]{G. Pettinari,}
\author[a,b]{F. Piacentini,}
\author[d]{\\M. Bersanelli,}
\author[d]{F. Cavaliere,}
\author[d]{C. Franceschet,}
\author[e]{M. Gervasi,}
\author[e]{\\A. Limonta,}
\author[d]{S. Mandelli,}
\author[d]{E. Manzan,}
\author[d]{A. Mennella,}
\author[e]{A. Passerini,}
\author[f]{\\E. Tommasi,}
\author[f]{A. Volpe,}
\author[e]{M. Zannoni}

\affiliation[a]{Dipartimento di Fisica, \emph{Sapienza} Universit\`a di Roma, P.le A. Moro 5, 00185 Roma, Italy}
\affiliation[b]{INFN sezione Roma1, P.le A. Moro 5, 00185 Roma, Italy}
\affiliation[c]{Istituto di Fotonica e Nanotecnologie -- CNR, Via Cineto Romano 42, 00156 Roma, Italy}
\affiliation[d]{Dipartimento di Fisica, Universit\`a di Milano \& INFN sezione Milano, Via Giovanni Celoria 16, 20133 Milano, Italy}
\affiliation[e]{Dipartimento di Fisica, Universit\`a di Milano Bicocca \& INFN sezione Milano Bicocca, Piazza della Scienza 3, 20126 Milano, Italy}
\affiliation[f]{ Agenzia Spaziale Italiana, Via del Politecnico, 00133 Roma, Italy}

\emailAdd{alessandro.paiella@roma1.infn.it}

\abstract{We have developed two arrays of lumped element kinetic inductance detectors working in the D--band, and optimised for the low radiative background conditions of a satellite mission aiming at precision measurements of the Cosmic Microwave Background (CMB) radiation. The first detector array is sensitive to the total power of the incoming radiation to which is coupled via single-mode waveguides and corrugated feed--horns, while the second is sensitive to the polarisation of the radiation thanks to orthomode transducers.

Here, we focus on the total power detector array, which is suitable, for instance, for precision measurements of unpolarised spectral distortions of the CMB, where detecting both polarisations provides a sensitivity advantage. We describe the optimisation of the array design, fabrication and packaging, the dark and optical characterisation, and the performance of the black--body calibrator used for the optical tests. We show that almost all the detectors of the array are photon-noise limited under the radiative background of a \SI{3.6}{K} black--body. This result, combined with the weak sensitivity to cosmic ray hits demonstrated with the OLIMPO flight, validates the idea of using lumped elements kinetic inductance detectors for precision, space--based CMB missions.
}


\begin{document}
\maketitle
\flushbottom

\section{Introduction}
\label{intro}
The measurements of primordial B--modes and spectral distortions of the Cosmic Microwave Background (CMB) radiation represent the challenges of the modern precision cosmology. 

The primordial B--modes \cite{B-modes} are generated by gravitational waves producing both curl--free and curly field of
polarisation vectors and they are predicted by inflation models. The detection of these modes would, therefore, bring an important confirmation to the inflation theory \cite{inflation}. The common methodology to deal with this measurement, adopted by the majority of the current experiments, is to have multi--band focal plane populated with thousands of independent detectors: \emph{multi--band} in order to efficiently remove the foreground signals and \emph{thousands of independent detectors} in order to increase the overall sensitivity of the instrument (which scales as the square root of the number of detectors). A clear example of this approach was the CORE satellite \cite{Delabrouille_2018,de_Bernardis_2018,Finelli_2018,Valentino_2018,Challinor_2018,Melin_2018,Zotti_2018,Burigana_2018,Natoli_2018,Remazeilles_2018}, proposed to but not funded by the European Space Agency, which should have had about \SI{2000}{detectors} in 19 bands ranging from 60 to \SI{600}{GHz}. In the past years, several experiments, both ground-based and balloon-borne as well as satellites, have been proposed and funded with the goal to measure the primordial B--modes: the ground-based BICEP 3 \cite{BICEP2} and Keck array \cite{BICEP}, POLARBEAR--2 \cite{polarbear,polarbear2}, QUBIC \cite{Qubic1,Qubic2,Qubic3}, Simons Observatory \cite{simons,simons2} and STRIP/LSPE \cite{STRIP,LSPE1,Addamo_2021}; the balloon-borne EBEX \cite{EBEX}, SPIDER \cite{SPIDER1,SPIDER2,spidercollaboration2021constraint} and SWIPE/LSPE \cite{LSPE1,LSPE3,LSPE2,Addamo_2021}; and the LiteBIRD satellite \cite{litebird,masashi,Lamagna_LiteBIRD}.   

The spectral distortions of the CMB are very small deviations from a pure Planck's black--body spectrum \cite{Kogut2019cmb,chluba2019new}. The current limit for the monopole component, measured by COBE/FIRAS \cite{Fixsen_1996} and TRIS \cite{Gervasi_2008}, indicates that these deviation are lower than 50 parts per million. Spectral distortions are predicted by the theory, and are due to non--equilibrium energy transfers between radiation and matter, happening before and after the recombination era: while those generated after the recombination, during the reionisation, are accessible with other observable such as the CMB anisotropies; the distortions sourced before the recombination are not accessible with techniques different from absolute spectroscopy and carry important information about the thermal history of the first stages of the universe evolution. The methodology of the measurement of the spectral distortions consists in the measurement of the difference between the spectrum of the CMB and a calibrator. It is clear, therefore, that the more the calibrator will have a pure Planck's spectrum, the more the measurement will be accurate. Moreover, very sensitive and broad--band detectors will be needed. An example is the PIXIE satellite \cite{Kogut_2011} proposed to NASA, which uses a cold Fourier transform spectrometer (FTS, cooled at \SI{2.725}{K}) and a calibrator with \SI{}{\micro\kelvin} accuracy. Recently, a ground--based experiment has been proposed and funded by PNRA (National Research Program in Antarctica) and PRIN (Projects of Relevant National Interest) in Italy as a pathfinder to measure the monopole spectral distortions: COSMO (COSmic Monopole Observer) will be installed at Dome--C in Antarctica and will be equipped with a cold differential FTS and multi--mode horn--coupled lumped element kinetic inductance detectors (LEKIDs). The experiment exploits the fast response of the KIDs to beat atmospheric emission fluctuations by means of fast sky dips, measuring simultaneously the sky monopole and the atmospheric emission. After validation of the method, COSMO will be transformed into a balloon--borne instrument, improving significantly its sensitivity to monopole spectral distortions. 

Anisotropic spectral distortions can be detected by exploiting the spatial variation across the observed sky region. Since an absolute determination of the CMB spectrum is not needed in this case, a differential spectrometer can be usefully coupled to a suitable large throughput optical system to measure the differential spectrum between different sky regions. An efficient spatial--spectral mapping towards specific sky patches can therefore be used to constrain the main parameters that quantify the amplitude of the distortion at the scales selected by the instrument. One such example is the observation of galaxy clusters through the Sunyaev--Zel'dovich (SZ) effect, a y--type distortion generated by inverse Compton scattering of CMB photons off electrons in the atmospheres of galaxy clusters (see \emph{e. g.} \cite{2002ARAA..40..643C}). For such measurements, the combination of high angular resolution and mapping speed over the whole relevant microwave bandwidth is a key requirement to detect the SZ signatures from arcminute--scale clusters by exploiting the typical spectral signature of the y--type distortion through broad--band microwave spectrometry. Examples of such an instrument design are the OLIMPO balloon--borne experiment \cite{Masi_2019,Coppolecchia2020b,Paiella_2020,Presta_2020} and the MILLIMETRON Space Observatory\footnote{http://www.millimetron.ru/en/} \cite{2009ExA....23..221W,2012RQE...54..557S} proposed to Roscosmos, both equipped with large aperture optics and a differential FTS.

The ASI/KIDS (Kinetic Inductance Detectors for Space) project is a R\&D activity, funded by the Italian Space Agency (ASI), devoted to the TRL (Technology Readiness Level) advancement of kinetic inductance detectors working in the D--band $\left[110;170\right]\SI{}{GHz}$ in view of future CMB satellite missions. In the past, an effort in this direction has been made by the SPACEKIDS project \cite{D'Addabbo,SPACEKIDS} and recently by \cite{catalano2020lekid}. The OLIMPO \cite{2013IAUS..288..208D,Masi_EPJ,Masi_2019,Paiella_2020} and BLAST--TNG \cite{BLAST1,BLAST2} balloon--borne experiments have already operated arrays of LEKIDs in the stratosphere. Kinetic inductance detectors \cite{Day} are superconducting resonators, where the radiation is detected by the breaking of Cooper pairs of the superconductor. In the lumped element configuration \cite{Doyle2008}, the absorption of the radiation happens in the inductor of the resonator and then the breaking of Cooper--pairs produces a change of the resonant frequency and the quality factor of the resonator, which can be read out by measuring the amplitude and phase of the bias signal transmitted past the resonator. The goal of the ASI/KIDS project concerns the development of two LEKID arrays: one sensitive to the total power and the other sensitive to the radiation polarisation, both coupled to the radiation via corrugate feed--horns \cite{mandelli}. In the polarisation sensitive array, the polarisation is separated in the two components by single--mode orthomode transducer.

In this paper, we will describe the design, the optimisation and the dark and optical characterisation of the 37--pixel LEKID array sensitive to the total power. The paper is organised as follows: section~\ref{sec:1} describes the design of the wafer holder optimised for efficient tiling of them in order to populate large focal planes; section~\ref{sec:2} describes the design of the LEKID array, the optical simulations of the receiver system and the fabrication process; section~\ref{sec:3} collects the results of the dark measurements on a prototype pixel of the final array; section~\ref{sec:4} collects the results of the optical characterisation of the 37--pixel array performed in a low radiative background and the description of the testbed; section~\ref{sec:5} discusses a possible application of this LEKID array on--board a space mission devoted to precision measurements of CMB spectral distortions; and in section~\ref{sec:6} we report the conclusion.

\section{Wafer holder design and feed--horn array}
\label{sec:1}

One of the goals of the ASI/KIDS project is to develop a receiver module which are easily tiled, in an efficient way, in large focal planes. The simplest solution would be a single large--diameter thin dielectric wafer hosting hundreds of LEKIDs, but this is not always possible neither from the manufacturing point of view nor from a mechanical point of view. The former because processing machines for large--diameter wafers are not widely available, the latter because, for a given thickness (dictated by the optical design, see section~\ref{sec:2}), the larger the diameter, the more fragile the wafer. This is particularly important for an array to be used in a space mission. 

\begin{figure}
\centering
\includegraphics[scale=0.4]{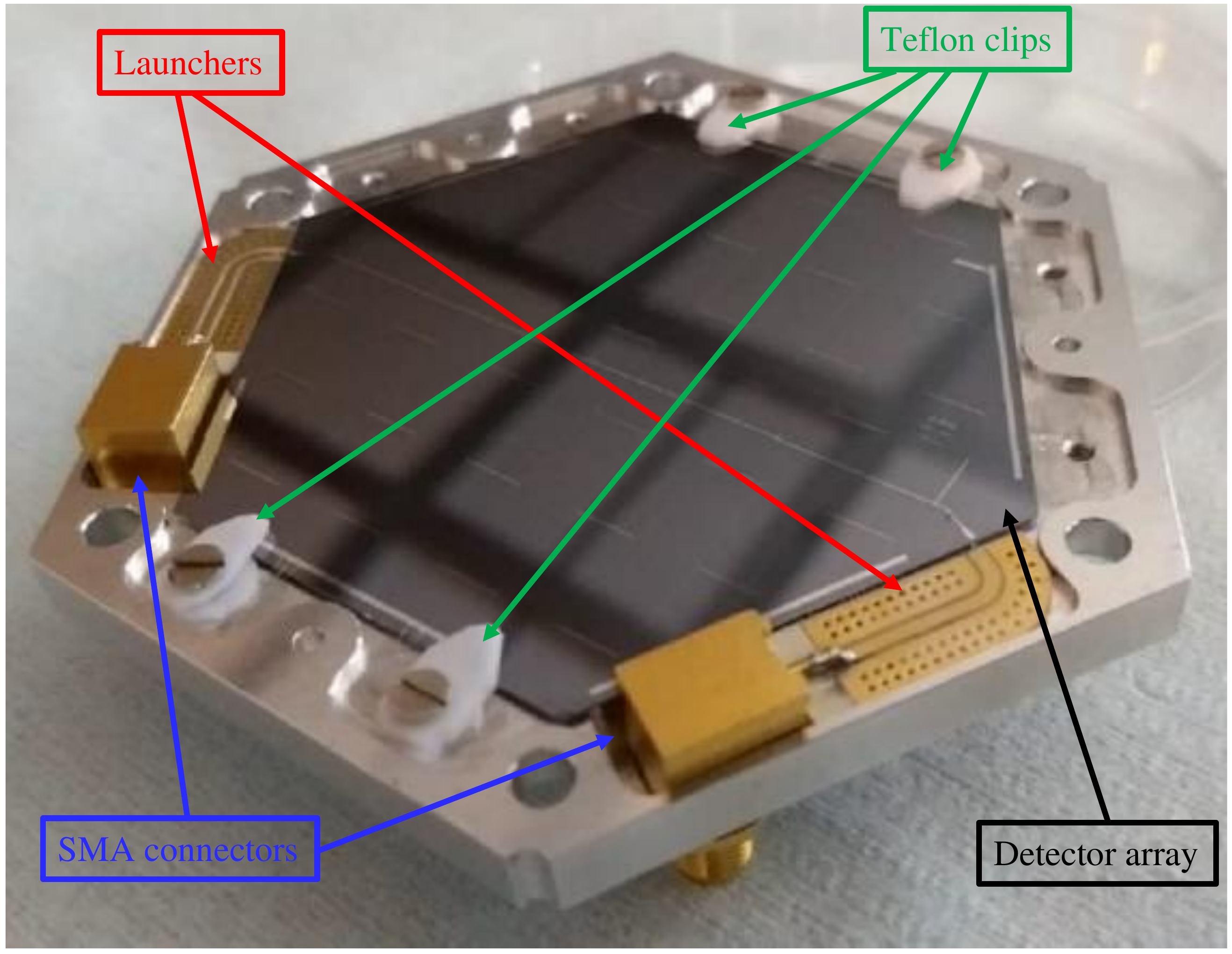}

\caption{\small Picture of the detector holder hosting the detector array, the SMA connectors, the launchers and the teflon clips which anchor the detector array to the holder.}
\label{fig:photo_Launcher}   
\end{figure}

For all these reasons, we prefer to use a silicon wafer with \SI{3}{inches} diameter, cutting it to a hexagon shape, and hosting it in a hexagon holder coupled to a hexagonal feed--horn array. The two SMA feedline connectors exit the rear side of the wafer, as shown in figure~\ref{fig:photo_Launcher}, so that it will be possible to populate large focal planes with an acceptable loss of pixels with respect to the case of a single larger wafer, by using a number of replicas of this modular hexagonal receiver unit (taking care to vary the resonant frequencies of the pixels of the different modules if they are all bias/readout with a single electronics).

\begin{figure}[!h]
\centering
\includegraphics[scale=0.45]{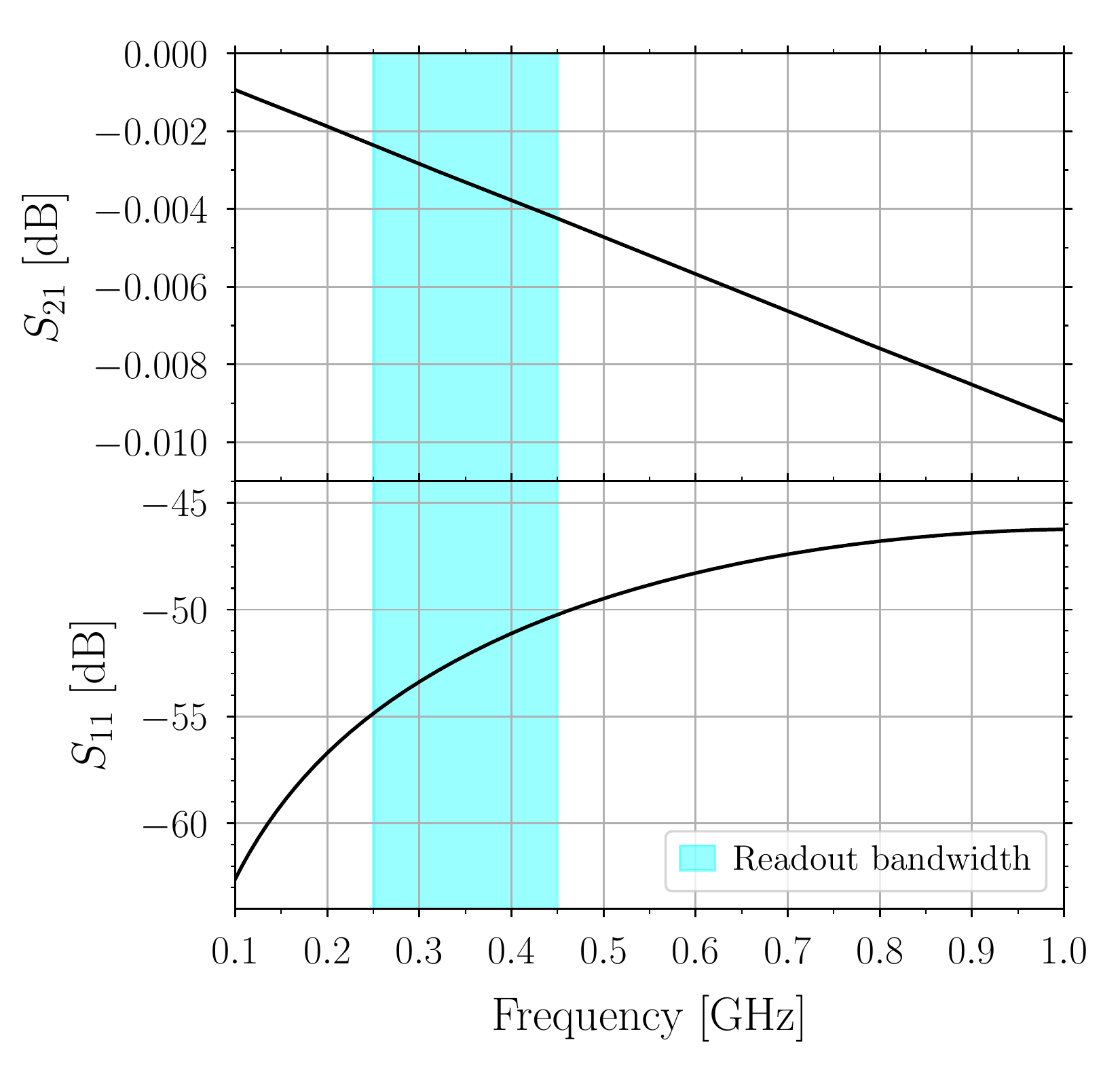}
\includegraphics[scale=0.45]{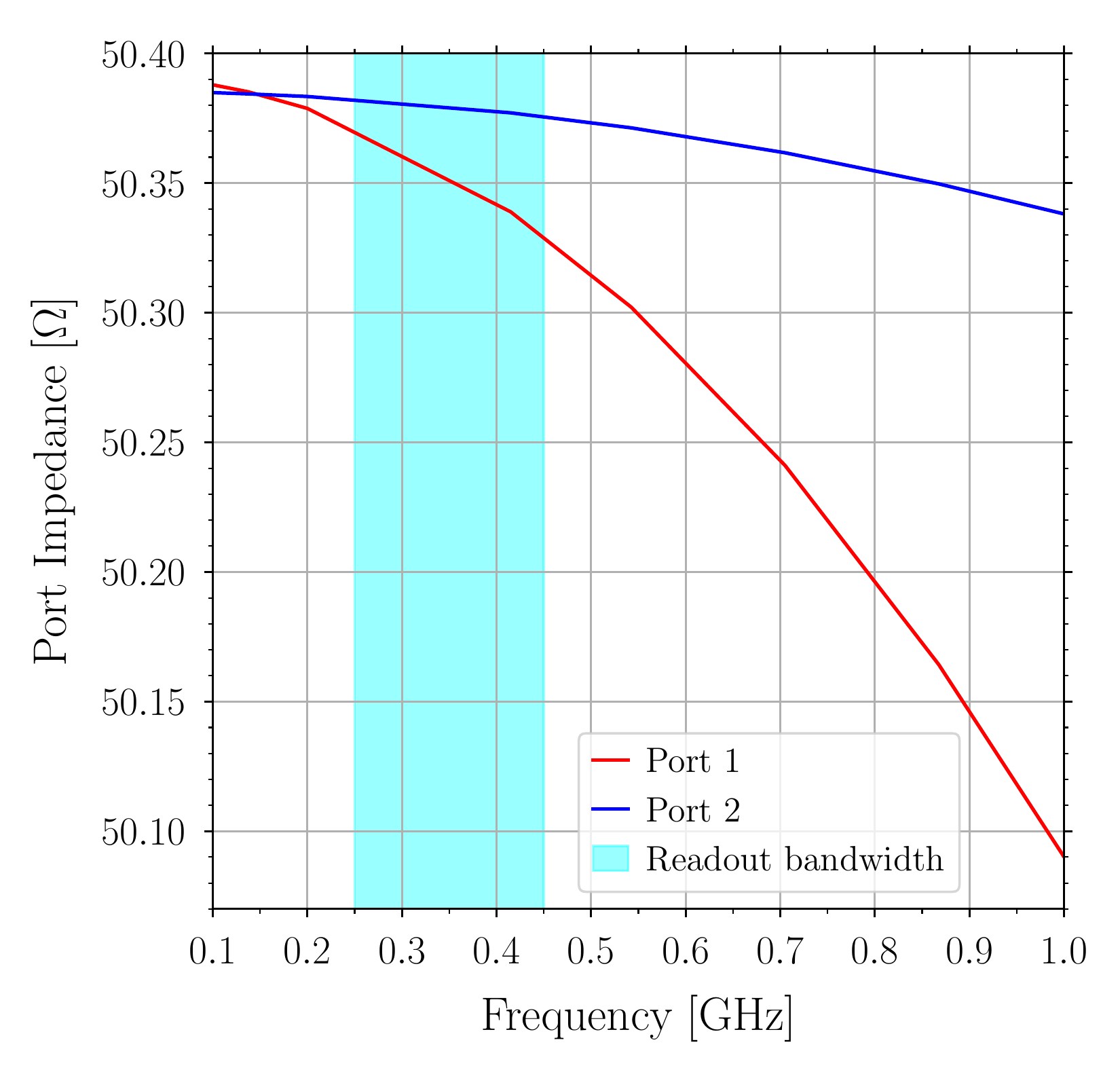}
\caption{\small \emph{Left panel}: SONNET simulation results for the launcher scattering parameters $S_{21}$ (\emph{top panel}) and $S_{11}$ (\emph{bottom panel}). \emph{Right panel}: SONNET simulation results for the launcher CPW impedance in the neighbourhood of the simulation ports, placed on the extremities of the CPW.}
\label{fig:simu_Launcher}     
\end{figure}

The bias signal is fed into the array feedline, from the SMA connectors, through the \emph{launchers} (see figure~\ref{fig:photo_Launcher}). They are coplanar waveguides (CPW), the impedance of which need to be matched to standard \SI{50}{\ohm} coaxial cables and interfaces in order to minimise losses.
The launchers are made on a Rogers RO4350B\footnote{https://rogerscorp.com/advanced-connectivity-solutions/ro4000-series-laminates/ro4350b-laminates} substrate \SI{0.5}{mm} thick, the CPW and the ground plane (on the opposite side of the substrate with respect to the CPW) are realised with chemical gold finish copper \SI{18}{\micro\metre} thick, and the vias are needed to allow the silver--epoxy glue to guarantee that the ground planes of the CPW are short--circuited. The \emph{left panel} of figure~\ref{fig:simu_Launcher} shows the SONNET\footnote{https://www.sonnetsoftware.com/} simulation results for the scattering parameters: we obtained a transmission scattering parameter $S_{21}\sim\SI{0}{dB}$, and a reflection scattering parameter $S_{11}<\SI{-50}{dB}$ across the readout bandwidth between \SI{250}{MHz} and \SI{450}{MHz}, see section~\ref{sec:2}. The \emph{right panel} of figure~\ref{fig:simu_Launcher} shows the SONNET simulation results for the impedance of the ports of the simulation: Port 1 is near the curved section and Port 2 is on the side of the straight section. As we can see, the impedance of the CPW results quite well matched to \SI{50}{\ohm}.

The corrugated feed--horn array, completed with the single--mode circular waveguide array, has been described and fully characterised in \cite{mandelli}, where it is possible to find all the information about the design, size and performance. The aperture of the feed--horns and the distance between them constrain the number of LEKIDs to be hosted on a \SI{3}{inches} substrate to 37.

\section{Detector array design, simulations and fabrication}
\label{sec:2}

For the design of the ASI/KIDS LEKID array, we started from the experience gained with the \SI{150}{GHz} array of the OLIMPO experiment \cite{Paiella_2017,Paiella_2019a} and with a R\&D activity of W--band LEKIDs \cite{Paiella_2016,Coppolecchia2020}, and remembering that they have to be operated in space. Therefore, since the background radiation experienced by detectors operated in space is lower than that experienced by detectors operated in the stratosphere, we can focus on the maximisation of the detector responsivity at the expense of the detector dynamics. 

Before going into the details of the design, we remind that the transmission scattering parameter, $S_{21}$, for a LEKID is given by \cite{Zmuidzinas}
\begin{equation}
S_{21}\left(\nu\right)\sim 1-\frac{Q_{tot}/Q_{c}}{1+2jQ_{tot}\displaystyle \frac{\nu-\nu_r}{\nu_r}}    
\label{eq:S_21}  
\end{equation}
with $Q_{tot}^{-1}=Q_{c}^{-1}+Q_{i}^{-1}$, where $\nu$ is the bias frequency, $\nu_r$ is the resonant frequency, $Q_{tot}$ is the total quality factor, $Q_{c}$ is the coupling quality factor and $Q_{i}$ is the internal quality factor. In the phase readout system, we take the phase $\varphi$ of equation~\ref{eq:S_21} and we define the electrical (or dark) responsivity as
\begin{equation}
\mathcal{R}_{elec,\varphi}=\frac{\delta\varphi}{\delta P_{elec}}=\frac{\delta\varphi}{\delta \nu_r}\frac{\delta\nu_r}{\delta N_{qp}}\frac{\delta N_{qp}}{\delta P_{elec}}=-\frac{\eta_{pb}\tau_{qp}}{\Delta_{0}}\frac{4Q_{tot}}{\nu_{r}}\frac{\delta\nu_{r}}{\delta N_{qp}}\;,
 \label{eq:resp_pha_elec_esp}
\end{equation}
where 
\begin{equation}
\delta P_{elec}=\frac{\Delta_0}{\eta_{pb}\tau_{qp}}\delta N_{qp} \notag
\end{equation}
is the electrical power producing the phase shift $\delta\varphi$, $\Delta_{0}=1.764k_{B}T_{c}$ is half Cooper--pair binding energy, $T_{c}$ is the critical temperature of the superconducting film, $\eta_{pb}\sim 0.57$ is the pair--breaking efficiency, $\tau_{qp}$ is the quasi--particle lifetime, and $\delta\nu_{r}/\delta N_{qp}$ is the shift of the resonant frequency with the number of quasi--particles, namely with temperature. The optical responsivity is
\begin{equation}
\mathcal{R}_{opt,\varphi}=\varepsilon\;\mathcal{R}_{elec,\varphi}
 \label{eq:optical_resp_defin}
\end{equation}
where $\varepsilon$ is the optical efficiency, which take into account the KID absorption efficiency $\eta$. Therefore we have
\begin{equation}
\mathcal{R}_{opt,\varphi}\propto\eta\;\mathcal{R}_{elec,\varphi}\propto\eta\frac{Q_{tot}\alpha}{V}\;,
 \label{eq:optical_resp_propto}
\end{equation}
where $\alpha=L_k/L$ is the ratio between the kinetic inductance and the total inductance ($L=L_k+L_g$ and $L_g$ is the geometric inductance) and $V$ is the detector absorber volume.

\begin{figure}[!h]
\centering
\includegraphics[scale=0.46]{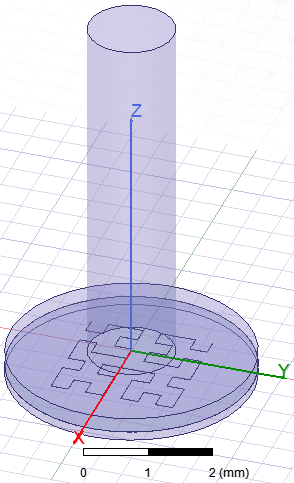}\hspace{1cm}
\includegraphics[scale=0.5]{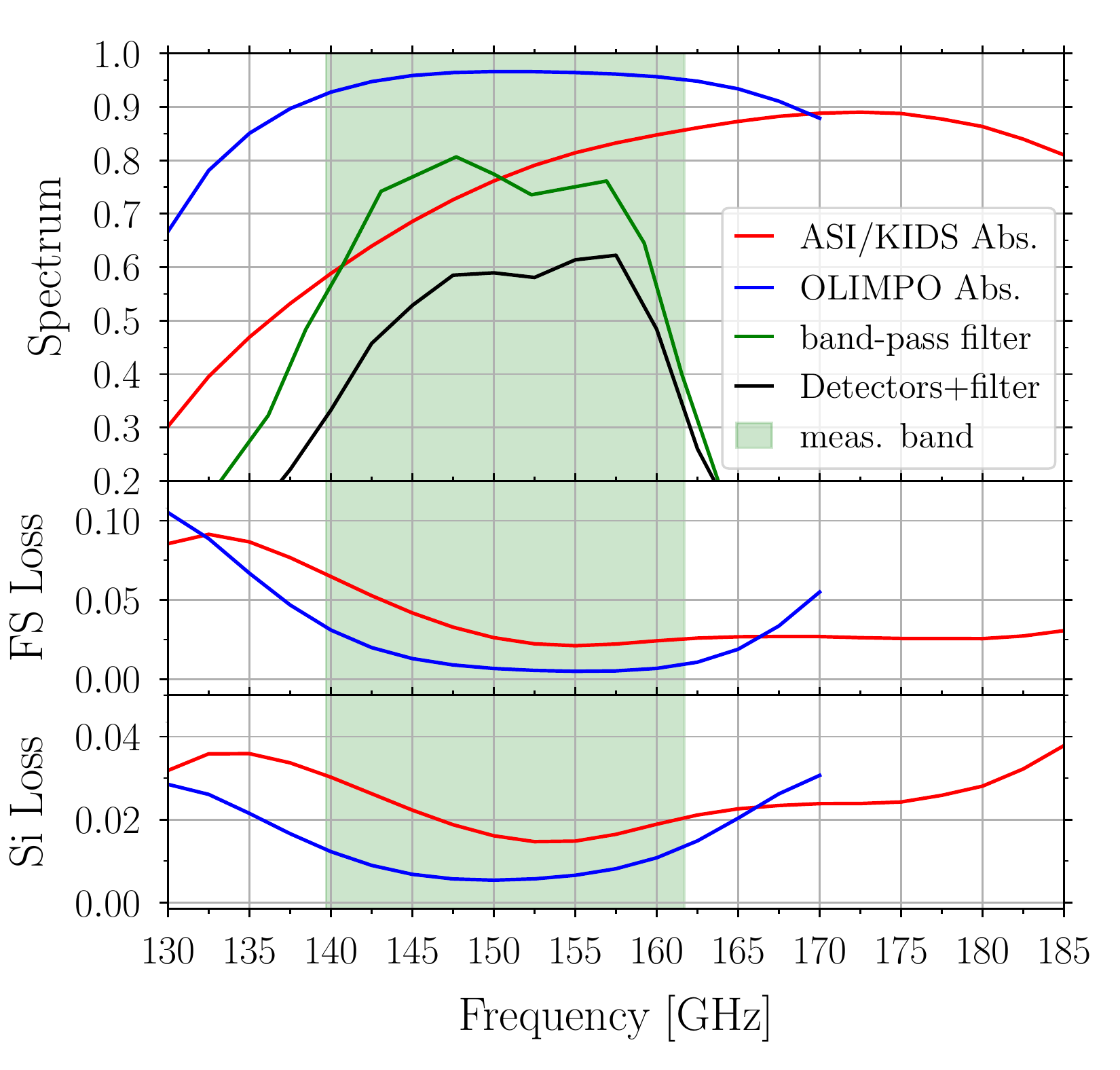}
\caption{\small \emph{Left panel}: HFSS design of the ASI/KIDS total power receiver from the circular waveguide to the backshort. \emph{Right panel}: HFSS results for the absorption efficiency compared to the band--pass filter transmission, and the expected optical efficiency spectrum of the detector$+$band--pass filter system (\emph{top panel}), and the free--space (\emph{central panel}) and silicon losses (\emph{bottom panel}) for the ASI/KIDS total power receiver, in \emph{red}, compared to those of the OLIMPO \SI{150}{GHz} receiver, in \emph{blue}.}
\label{fig:HFSS_simu}   
\end{figure}

The superconducting material is aluminum, which is widely used for LEKID arrays in the D--band \cite{NIKA,NIKA2,Heater1,Heater2,fasano}. The dielectric substrate is silicon \SI{135}{\micro\metre} thick, optimised for the OLIMPO \SI{150}{GHz} array \cite{Paiella_2019a}. The absorber of the detector, which is also the inductor of the resonator, has been optimised through Ansys HFSS\footnote{https://www.ansys.com/it-it/products/electronics/ansys-hfss} simulation, where we designed the receiver from the circular waveguide to the backshort, see the \emph{left panel} of figure~\ref{fig:HFSS_simu}. For the absorber geometry, we have chosen the 3$^{\rm rd}$ order Hilbert curve with characteristic length of \SI{285}{\micro\metre}, width \SI{2}{\micro\metre}, thickness \SI{25}{nm} and volume \SI{897.75}{\micro\metre^3}. Compared to the OLIMPO \SI{150}{GHz} absorber, which was a 4$^{th}$ order Hilbert curve with characteristic length of \SI{162}{\micro\metre}, width \SI{2}{\micro\metre}, thickness \SI{30}{nm} and volume \SI{2478.6}{\micro\metre^3}, the choice of the new geometry is motivated by the need to increase the responsivity which, according to equation~\ref{eq:optical_resp_propto}, scales as the inverse of the absorber volume ($V_{\rm ASI/KIDS}/V_{\rm OLIMPO}=0.36$), at the expense of the absorption efficiency ($\eta_{\rm ASI/KIDS}/\eta_{\rm OLIMPO}=0.80$) and the optical cross--talk, as shown in the \emph{right panel} of figure~\ref{fig:HFSS_simu}. This figure collects the HFSS results for the absorption efficiency and the losses\footnote{The losses are defined as the power which flows away through the side walls of the cylinders representing the substrate and the free--space between the waveguide exit and the absorber, therefore they can be used to set an upper limit to the optical cross--talk between neighbor absorbers.} of the new absorber (ASI/KIDS total power) compared to those of the OLIMPO \SI{150}{GHz} absorber. This plot also shows the transmission of the \SI{150}{GHz}--centered band--pass filter used for the optical characterisation, the expected optical efficiency spectrum of the detector$+$band--pass filter system and the corresponding measurement spectral band (used, in the following, for the computation of the background power and photon noise equivalent power, see figure~\ref{fig:meas_spectra} for the measurement of the spectral band).

Moreover, changing the thickness of the aluminum film means changing its critical temperature and kinetic inductance. Precisely, decreasing the aluminum thickness implies increasing both the critical temperature and the kinetic inductance. Increasing the critical temperature, for a fixed operating temperature, means increasing the kinetic inductance. Therefore decreasing the aluminum thickness implies to increase the kinetic inductance and then $\alpha$ (in our case $L_g\gg L_k\Rightarrow\alpha\approx L_k/L_g$): this is again, according to equation~\ref{eq:optical_resp_propto}, in the direction to increase the responsivity.

For the feedline architecture and the geometry of the capacitors we adopted the same as the OLIMPO array. The KID capacitors have been designed in order to have an expected internal quality factor, $Q_{i}$, of about \SI{50000}{}, at an operating temperature of \SI{185}{mK}, and to satisfy the \emph{lumped element condition}: the characteristic length of the circuit is much smaller than the operating wavelength of the circuit. This constrains the resonant frequencies in the range from \SI{250}{MHz} to \SI{450}{MHz}.

The electrical coupling between the KID and the feedline is obtained by means of capacitors, the capacitances of which have been chosen to constrain the coupling quality factors, $Q_{c}$. According to equation~\ref{eq:S_21} and \ref{eq:optical_resp_propto}, in order to maximise the responsivity, which scales as $Q_{tot}$, without degrading the resonator dip depth (defined as $S_{21}\left(\nu_r\right)$ from equation~\ref{eq:S_21}), which scales as $Q_{c}/\left(Q_{c}+Q_{i}\right)$, and without going in the condition $Q_{c}\gg Q_{i}\sim Q_{tot}$; we chose $Q_{c}\sim Q_{i}\sim\SI{50000}{}$ and therefore $Q_{tot}\sim Q_{i}/2\sim\SI{25000}{}$ and a resonator dip depth of about \SI{6}{dB}. This is justified by the low and stable background under which the detectors will work, and the expected small signals that the detectors would measure. In the case of OLIMPO, the \SI{150}{GHz} array was expected to work at an operating temperature of \SI{300}{mK}, with a modest variation of the background due to both the change of the elevation of the telescope boresight and the change of the measurement configuration, photometer and spectrometer. Therefore we chose to have a high resonator dip depth ($\sim \SI{15}{dB}$ at \SI{300}{mK} and $\sim\SI{30}{dB}$ at \SI{185}{mK}), by choosing $Q_{c}\ll Q_{i}$ and $Q_{c}\sim \SI{15000}{}$ at \SI{300}{mK}, at the expense of the responsivity, as described in \cite{Paiella_2019a}.   

The final array has been, therefore, populated with 37 pixels (this constraint comes out at the end of section~\ref{sec:1}) by changing the capacitance of the KID and coupling capacitors of each resonator, spacing their resonant frequencies of about \SI{5}{MHz} in the range $\left[250;450\right]\SI{}{MHz}$. 

\subsection{Fabrication and cutting procedure}

The fabrication of the LEKID arrays has been realised in a ISO-5/ISO-6 clean room ambient (CNR-IFN Rome, Italy) by thin-film fabrication and lithographic techniques. The detectors were realised on the surface of a double side polished, \SI{3}{inches}, intrinsic Si (100) wafer (${\rm resistivity}>\SI{10}{\kilo\ohm.cm}$). 

\begin{figure}[!h]
\centering
\includegraphics[scale=0.87]{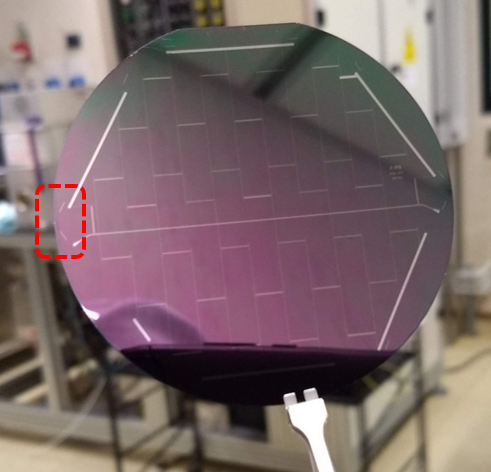}\hspace{-.0cm}
\includegraphics[scale=0.87]{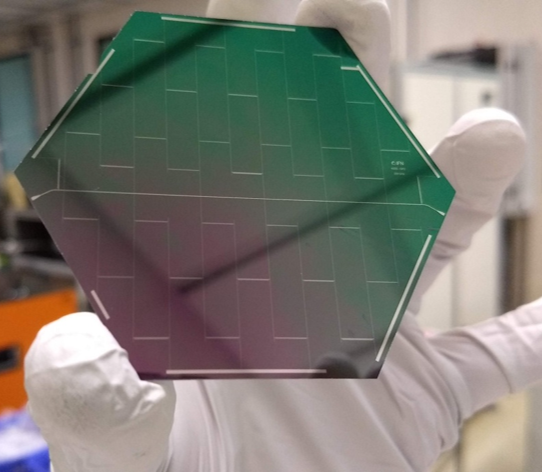}
\caption{\small The hexagonal shape cutting process. \emph{Left panel}: A series of markers along the vertices and sides of a regular hexagon profile, inscribed in the \SI{3}{inches} wafer, is drawn by EBL together with the realisation of the detector array; see area highlighted by the dashed rectangle. \emph{Right panel}: Picture of the hexagonal--shaped wafer after the cutting process (and before the protective PMMA layer removal).}
\label{fig:wafer_cutting}   
\end{figure}

\paragraph{Initial cleaning.} The wafer has been cleaned to remove native oxides from its surface just before the fabrication process. The cleaning consisted in a deep immersion in a HF (hydrogen fluoride) solution (\SI{5}{\%}, \SI{60}{\s}), followed by deep immersion in deionised (DI) water (\SI{20}{\s}) and blow dry in nitrogen flux (\SI{60}{\s}).
\paragraph{Resist deposition.} One face of the wafer was covered with a \SI{500}{nm} thick, uniform layer of polymethyl methacrylate (PMMA 600K, \SI{6}{\%} diluted in ethyl lactate) electron--resist. The resist has been spun at ambient temperature (\SI{20}{\degree C}) on the wafer by spinning it at \SI{3000}{rpm} for \SI{90}{\s}, and than baked on a hot plate at \SI{170}{\degree C} for \SI{300}{\s} in order to dry off solvent excesses. 
\paragraph{Litography.} The desired KID array layout was transferred by direct--writing electron beam lithography (EBL, working at \SI{100}{\kV}, clearance dose of \SI{550}{\micro C/cm^2}) on the PMMA layer: the electron irradiated area undergo a chemical modification (polymeric chain scission) that makes the exposed PMMA soluble in a proper solution, \emph{i. e.} a (1:1) solution of methyl isobutyl ketone (MIBK) and isopropyl alcohol (IPA). 
\paragraph{Cleaning.} The PMMA patterned wafer was subject to a mild O$_2$--plasma treatment (radio frequency of \SI{33}{\W}, bias voltage of \SI{90}{\V}, O$_2$ flux of \SI{196}{sccm}, base pressure of \SI{100}{mTorr}, \SI{30}{\s}) to clean up the Si uncovered area from PMMA residuals and to provide an undercut of the PMMA profile, which favours the lift-off process. 
\paragraph{Metal deposition -- KID array.} A second HF cleaning bath (\SI{2.5}{\%}, \SI{10}{\s}) was then performed just before placing the patterned wafer in the vacuum chamber of an e--gun metal deposition system, in order to control the native oxygen formation on the uncovered Si surface. There, a uniform layer of pure Al was deposited on the patterned surface by using a high metal deposition rate (\SI{1}{\nm/\s}, base pressure of \SI{1e-7}{mbar}) to guarantee a good film quality. The deposition rate and thickness have been monitored in--situ by a quartz micro--balance, whilst the final metal thickness has been then checked ex--situ with a mechanical profilometer.
\paragraph{Lift--off.} The wafer was immersed in a hot acetone bath (\SI{56}{\degree C}) and ultrasonic waves were applied to dissolve the residual PMMA under the Al and remove the metal excess to complete the lift--off process. Subsequent clean acetone baths were used to prevent sample contamination from small metal debris released in the solution during the lift--off. The lift-off process progress was controlled by optical microscope inspection with a 100$\times$ objective. The sample was finally cleaned by a deep immersion in IPA for \SI{60}{\s}, followed by a blow dry in nitrogen flux, leaving behind a clean Si wafer patterned with the Al film only in the desired area (the KID array layout).
\paragraph{Metal deposition -- Backshort.} The back face of the Si wafer (the face opposite to that where the detectors are realised) was uniformly covered with a \SI{200}{nm} thick layer of Al, which acts as backshort for the incoming radiation, being the front side of the wafer protected with a PMMA layer for all the duration of this fabrication step and up to its removal in acetone at the end of the whole fabrication process. The back--side metallisation has been performed by means of a DC sputter system (working at a power of \SI{430}{\W} and base pressure of  \SI{2e-6}{mbar}). A constant deposition rate (\SI{1.3}{\nm/\s}) have been maintained during the deposition and the final metal thickness has been checked ex--situ with a mechanical profilometer. 
\paragraph{Wafer cutting.} The patterned wafer was cut into a hexagonal shape by means of a manual diamond--tip scribe machine. The alignment between the KID array layout and the hexagonal shape was achieved by a series of markers (visible under the optical microscope of the scribe machine and used as guidelines for the cut) that have been realised together with the realisation of the detectors. The wafer has been then hand--cleaved along the hexagonal perimeter groves performed with the diamond tip. Although the hexagonal shape does not match the easy direction of cleaving of the Si (100) lattice, the cutting process we used allowed us to realise nearly perfect hexagonal shape without the need of a cumbersome wafer dicing system, see figure~\ref{fig:wafer_cutting}.

The realised detector array was finally packaged in its holder by keeping the wafer in place with four, \SI{1}{mm} thick, teflon clips. The electrical contact between the detector feedline and ground plane with the coaxial connectors on the holder was guaranteed by a series of \SI{25}{\micro\metre} thick Al wire bondings performed with a semiautomatic wedge wire bonder, see figure~\ref{fig:photo_Launcher}.

\section{Dark measurements}
\label{sec:3}

The dark measurements have been performed on a 11--pixel prototype populated with resonators with different values of $Q_{i}$ and $Q_{c}$. The detector holder has been closed with an aluminum cover placed at the same distance from the detector array at which the exits of the waveguides of the feed--horn array have to be placed. The detector array has been cooled to \SI{185}{mK} with a dilution refrigerator \cite{crio}, and the bias/readout cold and warm electronics are the same described and optimised in \cite{Paiella_2019b}.
The bias/readout electronics is a FPGA--based system composed of a ROACH--2 board\footnote{\url{https://casper.berkeley.edu/wiki/ROACH2}}, and a MUSIC DAC/ADC board\footnote{\url{https://casper.berkeley.edu/wiki/MUSIC Readout}}. The FPGA firmware has been
developed by Arizona State University \cite{Gordon2016}. This electronics can generate up to 1000 tones over a \SI{512}{MHz} bandwidth (precisely in the $\left[-256;256\right]\SI{}{MHz}$ range), with a sampling rate up to \SI{1}{kHz}. In order to feed resonators with resonant frequencies in the range $\left[250;450\right]\SI{}{MHz}$ the electronics has been equipped with an up-- and down--conversion system of microwave components.

The critical temperature of the \SI{25}{nm} thick Al film has been measured by monitoring with the vector network analyzer (VNA) at which temperature the $S_{21}$ parameter suddenly grows, which corresponds to the drop of the feedline resistance due to the superconductivity. We measured $T_{c}\sim \SI{1.43}{K}$, a sheet resistance before the transition of $R_{\Box}\sim\SI{1.04}{\ohm/\Box}$ and a kinetic inductance of $L_{k}\sim\SI{1.00}{pH/\Box}$.  

The following analysis regards the pixel which has been chosen to be replicated in the final array, namely the one with the quality factors closest to the desired $Q_{c}\sim Q_{i}\sim\SI{50000}{}$. The resonant frequency, $\nu_{r}$, the quality factors, the resonator dip depth and the quasi--particle lifetime, $\tau_{qp}$, measured for this pixel, are collected in table~\ref{tab:1}.

\begin{table}[!h]
\centering
\caption{\small Electrical parameters for the best resonator.}
\vspace{2mm}
\fontsize{10pt}{15pt}\selectfont{
\begin{tabular}{cc}
\hline
\hline
Parameter & Value  \\
\hline
$\nu_{r}$ $\left[\SI{}{MHz}\right]$ & 389.5  \\
$Q_{c}$ & \SI{66000}{}  \\
$Q_{tot}$ & \SI{25000}{} \\
$Q_{i}$ & \SI{43000}{} \\
resonator dip depth $\left[\SI{}{dB}\right]$& 4.4\\
$\tau_{qp}$ $\left[\SI{}{\micro\second}\right]$& \SI{40}{}\\
\hline\hline
\end{tabular}}
\phantomsection\label{tab:1} 
\end{table}

\begin{figure}[!h]
\centering
\includegraphics[scale=0.45]{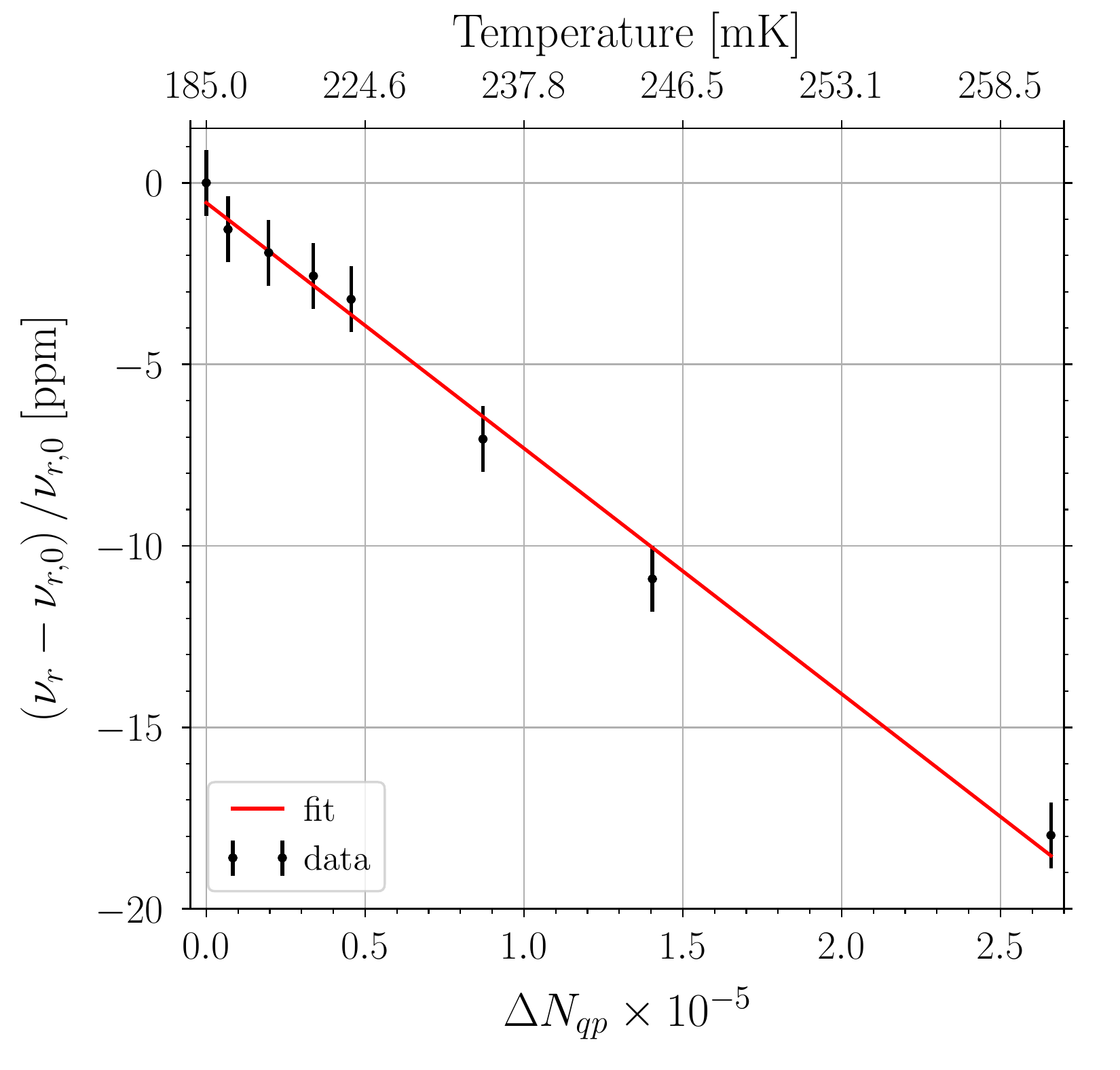}
\includegraphics[scale=0.45]{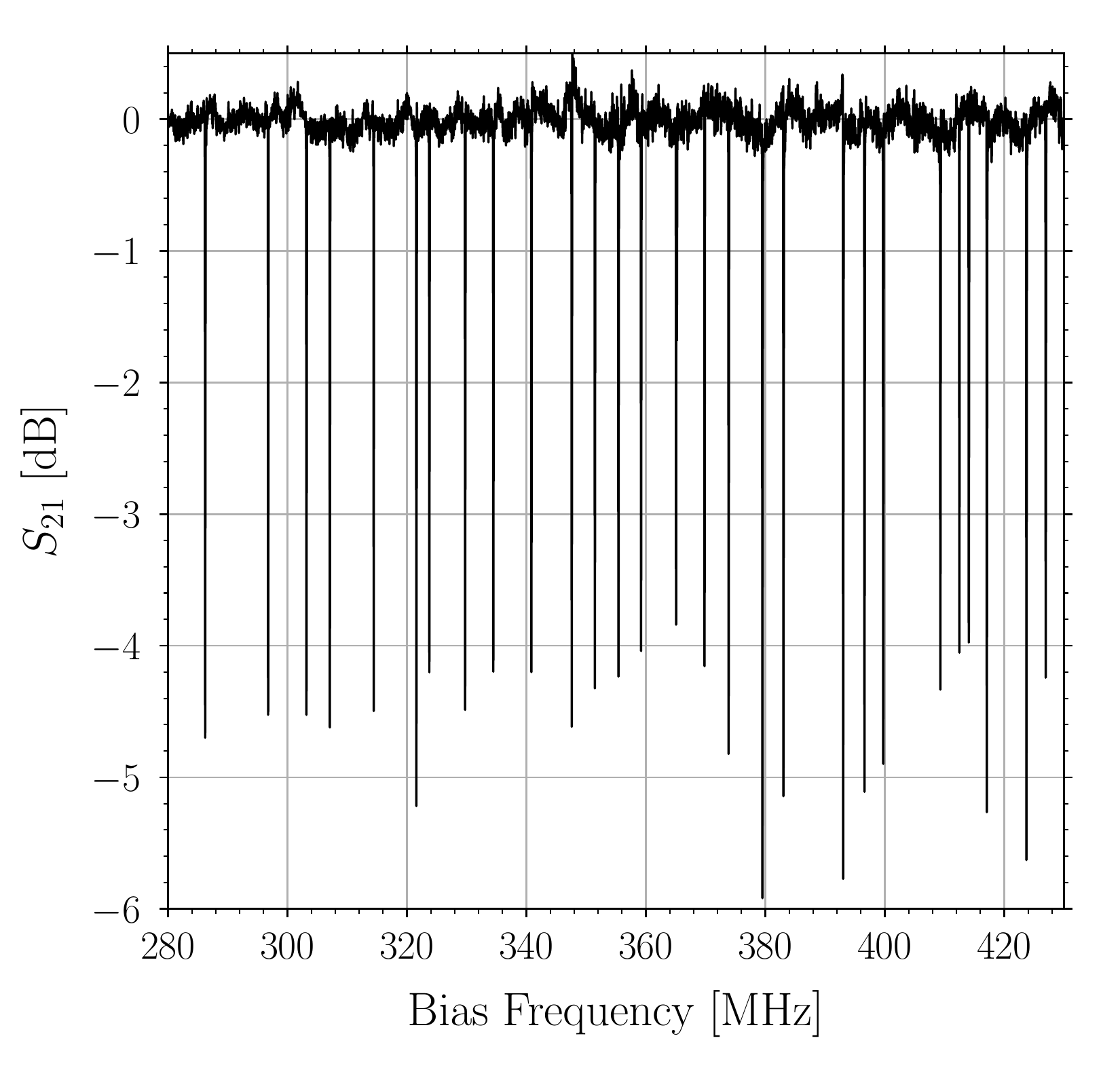}
\caption{\small \emph{Left panel}: Trends of the normalised resonant frequency shift with the number of quasi--particles. The \emph{black dots with error bars} are the measured data and the \emph{red line} is the linear fit. \emph{Right panel}: Amplitude of the $S_{21}$ parameter measured for the ASI/KIDS total power array. We found 28/37 active resonators: each \emph{vertical line} is a resonance}.
\label{fig:elec_resp_S21tot}   
\end{figure}

The electrical responsivity in phase, $\mathcal{R}_{\varphi}$, can be estimated using equation~\ref{eq:resp_pha_elec_esp} where $\delta\nu_{r}/\delta N_{qp}$ is the shape of the straight line defined by the resonant frequency shift \emph{versus} the number of quasi--particle increase, measured through a sweep of the operating temperature of the detectors, performed between \SI{185}{mK} and \SI{260}{mK}, see the \emph{left panel} of figure~\ref{fig:elec_resp_S21tot}. Therefore, using the measured values, we found
\begin{equation}
\mathcal{R}_{elec,\varphi}\sim\SI{1.2e13}{W^{-1}}\;.
 \label{eq:2}
\end{equation}
In order to obtain the electrical noise equivalent power (NEP) in phase, we have measured the dark noise spectrum: $\mathcal{N}_{dark,\varphi}\sim\SI{2.7e-5}{\sqrt{\rm Hz^{-1}}}$. Therefore, the electrical NEP in phase is ${\rm NEP}_{elec,\varphi}\sim\SI{2.3e-18}{W/\sqrt{\rm Hz}}$.

As a comparison, and with the aim to show that we went in the right direction, table~\ref{tab:2} collects all these results and those obtained for the OLIMPO \SI{150}{GHz} array characterised at the same operating temperature. As discussed before, the ASI/KIDS LEKIDs are made of a thinner aluminum film and they have a smaller absorber area with respect to the OLIMPO LEKIDs. Therefore the ASI/KIDS LEKIDs have a smaller volume, a greater $T_{c}$, a greater $\tau_{qp}$, a smaller $Q_{tot}$, and a greater $\alpha$ than OLIMPO KIDs. According to equation~\ref{eq:resp_pha_elec_esp} and \ref{eq:optical_resp_propto}, the combination of these effects goes in the direction of increasing the (electrical) responsivity.

\begin{table}[!h]
\centering
\caption{\small Comparison of the dark measurements between the ASI/KIDS total power pixel and the OLIMPO \SI{150}{GHz} array.}
\vspace{2mm}
\fontsize{10pt}{18pt}\selectfont{
\begin{tabular}{ccc}
\hline
\hline
Parameter & ASI/KIDS & OLIMPO \cite{Paiella_2019a}  \\
\hline
$T_{c}$ $\left[\SI{}{K}\right]$ &1.43 &1.31\\
Volume $\left[\SI{}{\micro\metre^{3}}\right]$ & 897.75 &2478.6  \\
$Q_{c}$ & \SI{66000}{}  & \SI{29400}{}\\
$Q_{tot}$ & \SI{25000}{} & \SI{28000}{} \\
resonator dip depth $\left[\SI{}{dB}\right]$& 4.4 &30\\
$\tau_{qp}$ $\left[\SI{}{\micro\second}\right]$& \SI{40}{}& \SI{30}{}\\
$\mathcal{R}_{elec,\varphi}$ $\left[\SI{}{W^{-1}}\right]$ &\SI{1.2e13}{} &\SI{2.3e12}{}\\
$\mathcal{N}_{dark,\varphi}$ $\left[\SI{}{\sqrt{\rm Hz^{-1}}}\right]$ &\SI{2.7e-5}{} &\SI{6.1e-5}{}\\
${\rm NEP}_{elec,\varphi}$ $\left[\SI{}{W/\sqrt{\rm Hz}}\right]$ &\SI{2.3e-18}{} &\SI{45e-18}{}\\
\hline
\hline
\end{tabular}}
\label{tab:2}  
\end{table}

Finally, we measured the amplitude of the $S_{21}$ parameter of the final array, shown in the \emph{right panel} of figure~\ref{fig:elec_resp_S21tot} from which we find 28/37 active resonators (referring to the figure, each depth in the transmission $S_{21}$ parameter, namely each \emph{vertical line}, corresponds to a resonance and therefore to an active resonator), namely a yield of 76\% (4 pixels were lost during the wafer cutting procedure). Moreover, the mean resonator dip depth is about \SI{4.4}{dB} as well as the quality factors are compatible with the ones collected in table~\ref{tab:1}.

\section{Optical characterisation}
\label{sec:4}

As we said before, the optical characterisation of the receiver (detector array $+$ feed--horn and waveguide array) has to be performed in a space--like environment, under a low radiative background condition. For this reason, the optical tests have been performed in a dark dilution cryostat, with a base temperature of \SI{185}{mK}, able to host a black--body calibrator on the \SI{3.6}{K} stage. We choose this temperature because it is representative of the radiative background of an experiment measuring the CMB from a satellite platform with a cryogenically cooled telescope.

\begin{figure}[!h]
\centering
\includegraphics[scale=0.104]{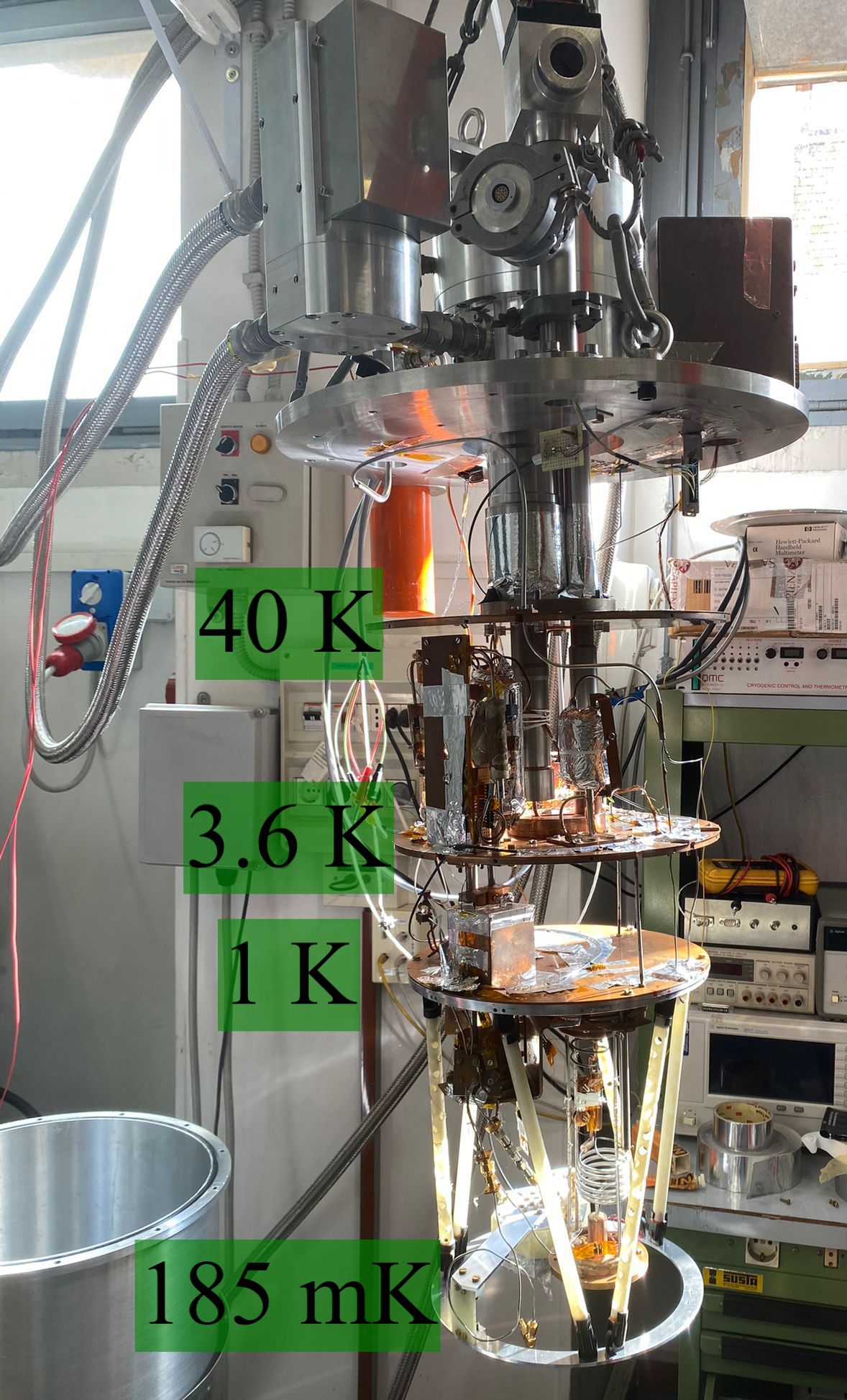}$\:$
\includegraphics[scale=0.099]{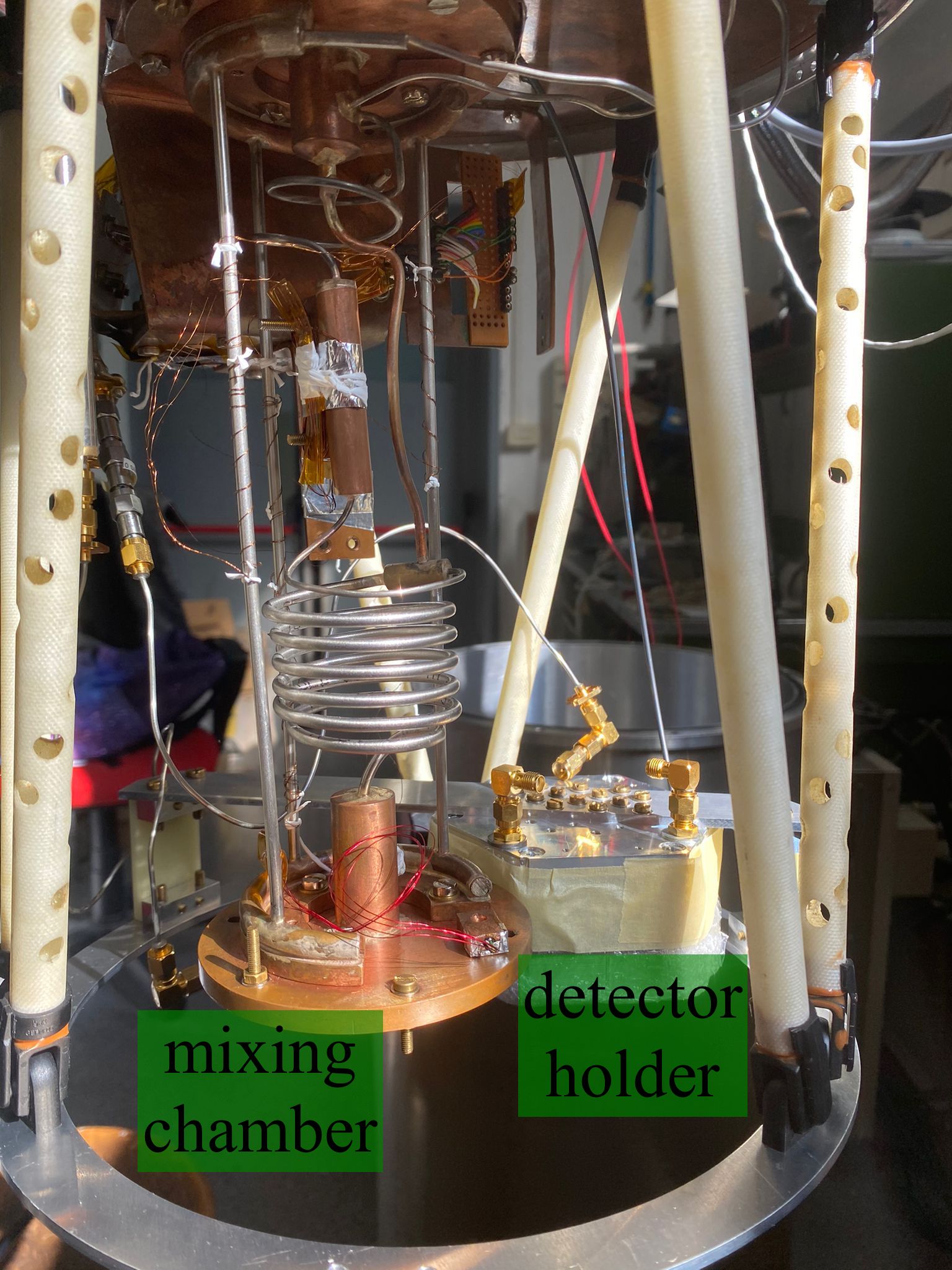}$\:$
\includegraphics[scale=0.099]{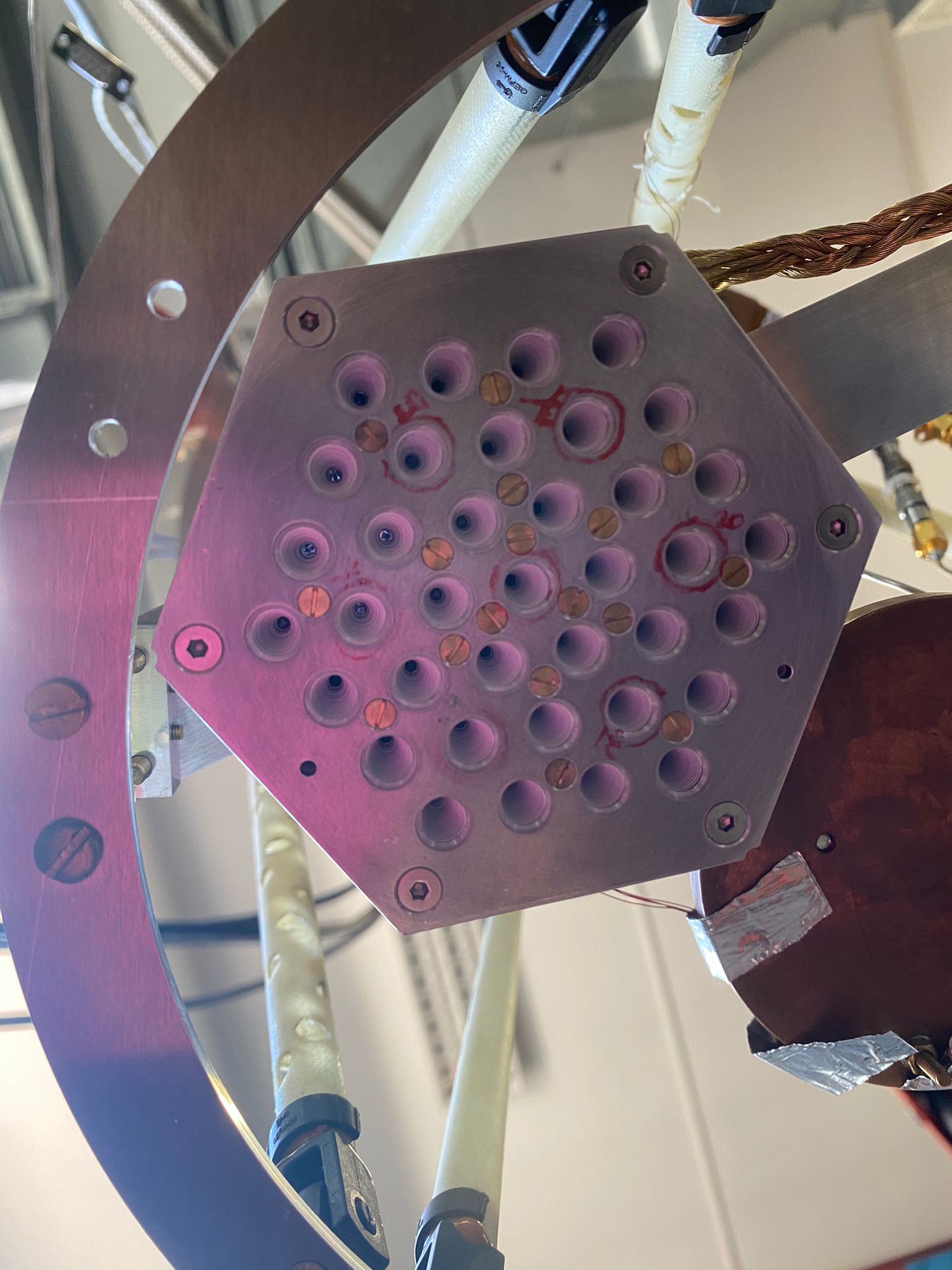}
\caption{\small Pictures of the receiver mounted in the cryostat by means of the aluminum--fiberglass support, and in thermal contact with the mixing chamber. \emph{Left panel}: Overall picture of the cryostat. \emph{Central panel}: Zoom of the cryostat cooldest stage and the detector holder. \emph{Right panel}: Bottom view of the feed-horn array mounted in the cryostat.}
\label{fig:photo_cryo}   
\end{figure}

As shown in figure~\ref{fig:photo_cryo}, the receiver has been integrated in the testbed using a custom aluminum--fiberglass support, in such a way that the feed--horn apertures are at a distance of few centimeters from the emitting surface. The aluminum--fiberglass support is composed of six fiberglass rods, anchored on the \SI{1}{K} stage, supporting and thermally insulating an aluminum ring which hosts the receiver, sustained by an aluminum bar, in turn, supported and thermally insulated by two fiberglass slabs. The detector holder is in thermal contact with the mixing chamber by means of a gold--plated copper strap. Along the optical path between the feed--horn apertures and the black--body calibrator, a \SI{240}{GHz} low--pass filter at \SI{3.6}{K}, and a \SI{150}{GHz}--centered band--pass filters at \SI{1}{K} have been placed to select the interesting band and to reduce the out--of--band background power on the detectors, see the \emph{left panel} of figure~\ref{fig:photo_BB}.

\subsection{Black--body calibrator}

\begin{figure}[!h]
\centering
\includegraphics[scale=0.45]{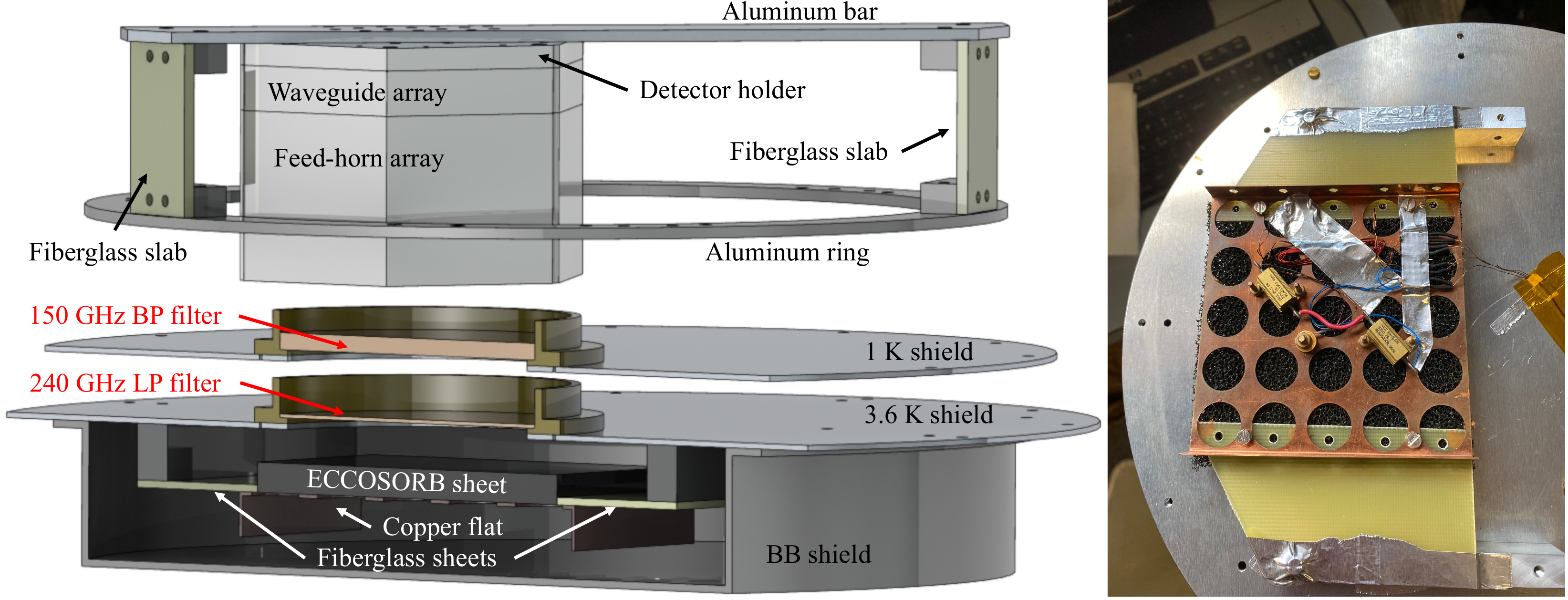}
\caption{\small \emph{Left panel}: Rendering showing the black--body calibrator and its radiation shield, the optical system, the detector holder and feed--horn array, and part of the aluminum-fiberglass support. \emph{Right panel}: Picture of the black--body source developed to load the KID array with a controlled, variable radiative background. The emitting surface is an ECCOSORB HR10 sheet, glued to a lightened copper flat, supported and thermally insulated by two fiberglass sheets. The copper flat is heated by two resistors, and its temperature is sensed by a Si diode thermometer.}
\label{fig:photo_BB}      
\end{figure}

We have developed and fabricated a custom device able to produce a controlled radiative background, similar to the one expected for space operation of the KIDs. It is composed of an ECCOSORB HR10\footnote{https://www.laird.com/products/microwave-absorbers/microwave-absorbing-foams/eccosorb-hr} sheet \SI{8}{mm} thick, supported by a \SI{0.5}{mm} thick, lightened copper flat. This is supported and thermally insulated by two thin (\SI{1.5}{mm} thick, \SI{100}{mm} wide, \SI{4}{mm} long) fiberglass sheets. The copper flat is heated by two \SI{330}{\ohm} metal--case resistors in series, and its temperature is sensed by a Si diode thermometer, see the \emph{left panel} of figure~\ref{fig:photo_BB}. This calibrator in anchored on the external surface of the case back of the \SI{3.6}{K} shield, and it is screened from the \SI{40}{K} background of the pulse tube first stage by means of an aluminum shield, see the \emph{left panel} of figure~\ref{fig:photo_BB}.

We have chosen an ECCOSORB HR10 foam as the microwave absorber, due to its high emissivity ($>90\%$ at \SI{140}{GHz}) and small mass and heat capacity. The drawback is its small heat conductivity. The system has been designed for a relatively fast time constant (tens of seconds) for the copper flat. The ECCOSORB HR10 foam will follow with a longer time constant, due to its internal heat conductivity. 

\begin{figure}[]
\centering
\includegraphics[scale=0.45]{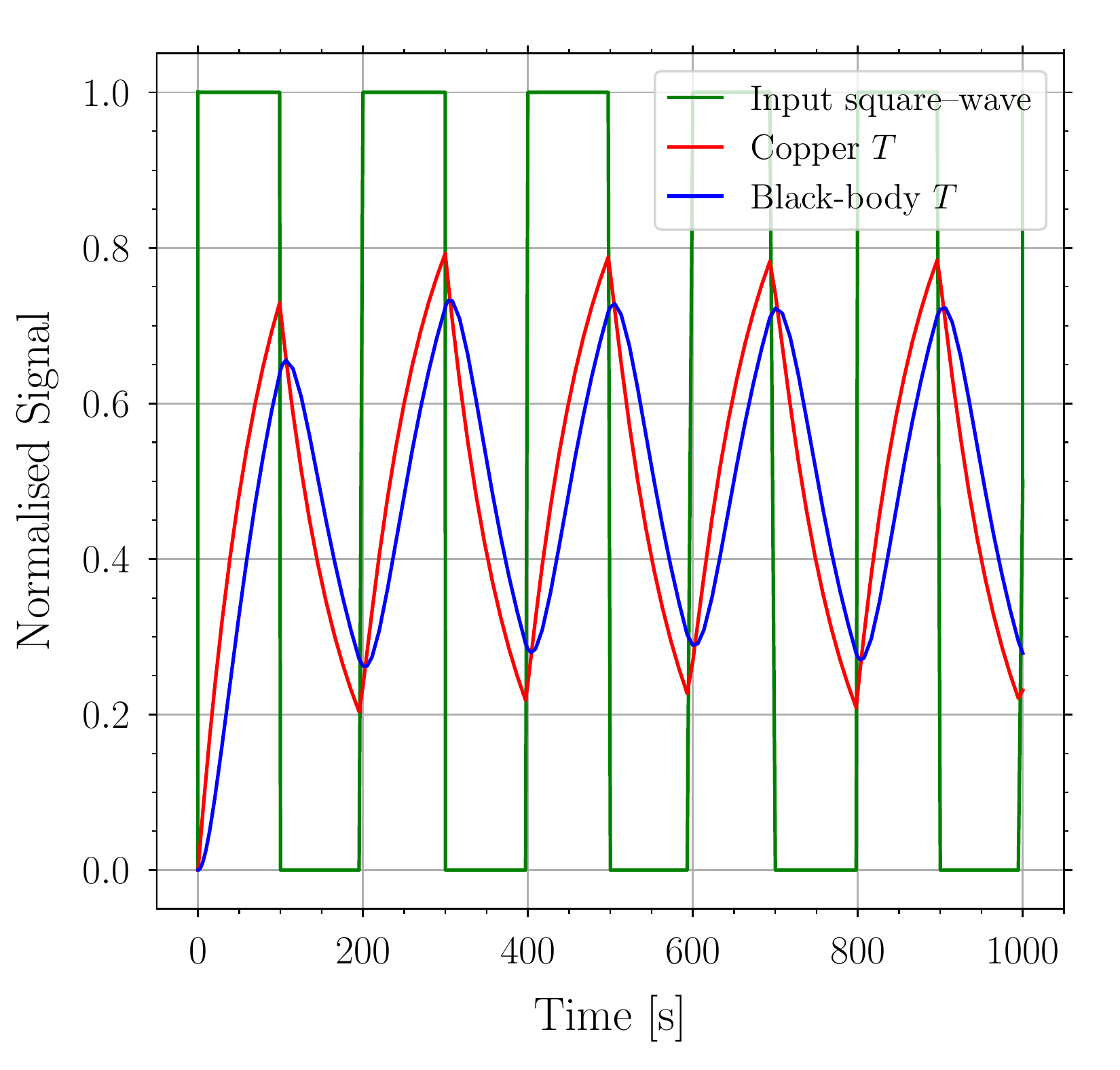}
\includegraphics[scale=0.45]{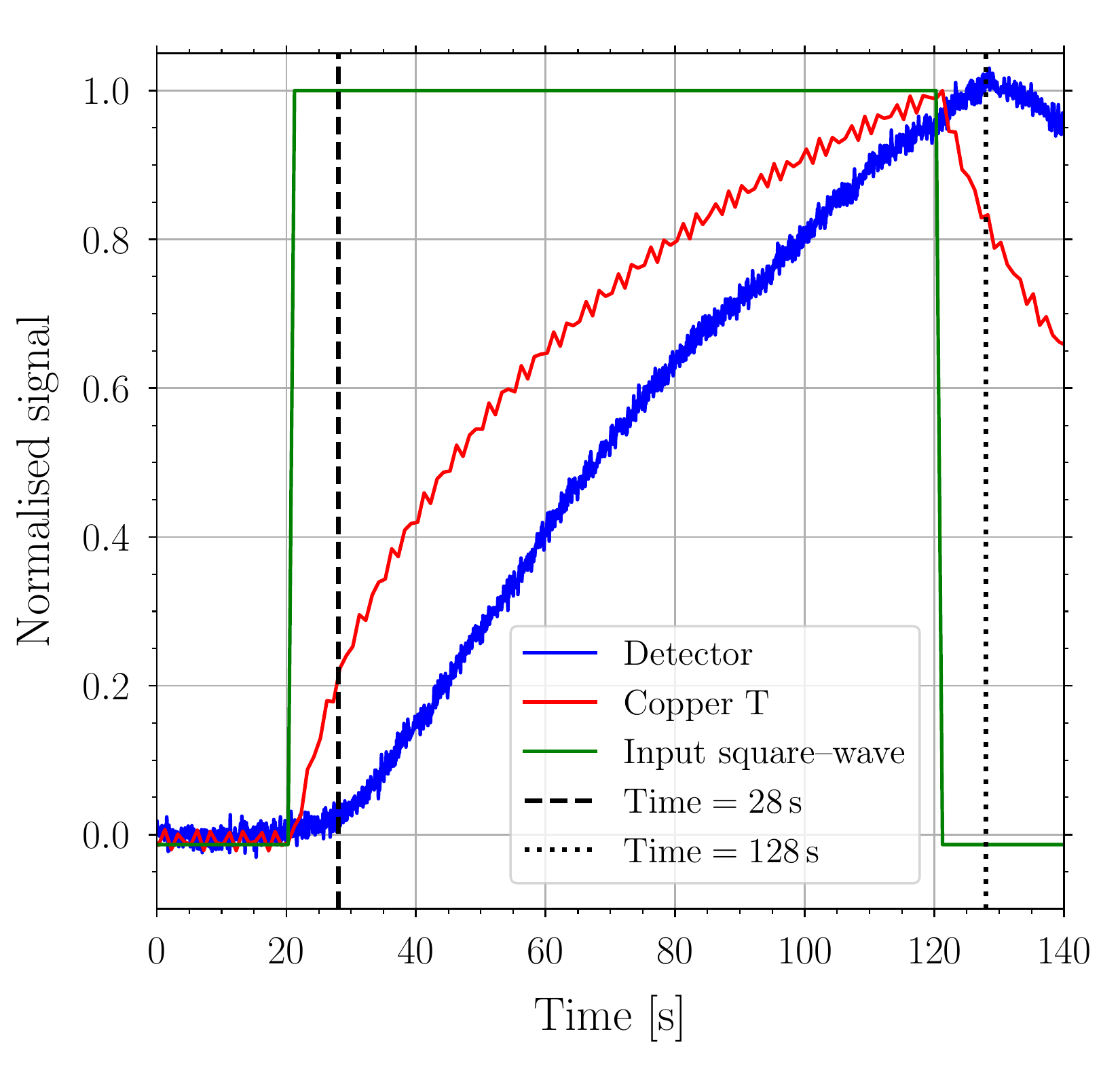}
\caption{\small \emph{Left panel}: Expected thermal response of the copper (\emph{red}) and black--body foam (\emph{blue}) under a \SI{200}{s} period square--wave excitation power dissipated in the resistors (\emph{green}). The copper temperature is normalised to its asymptotic value. The black--body foam temperature variation $\Delta T_{F}$ is smaller than the copper temperature variation $\Delta T_{Cu}$, due to its longer time constant. \emph{Right panel}: Measured thermal response of the copper (\emph{red}) and detector (\emph{blue}), normalised to their maximum values, under a \SI{200}{s} period square--wave excitation power dissipated in the resistors (\emph{green}). A semi--period is shown with the measurement of the delay between the maximum temperature of the copper plate and the maximum signal of the detector: $\sim\SI{8}{s}$.}
\label{fig:model_BB}      
\end{figure}

A simplified thermal model has been used, based on a two thermal time constant system, providing the expected behaviour shown in the \emph{left panel} of figure~\ref{fig:model_BB}. When the excitation is switched on, directly dissipating power on the copper plate, the temperature of the copper plate increases, with a time constant defined by its heat capacity and the conductivity of the fiberglass supports. In the example of the \emph{left panel} of figure~\ref{fig:model_BB}, this time constant is $\tau_{1} = $ \SI{75}{s} (consistent with direct measurements, see below). The temperature of the ECCOSORB HR10 foam increases with its own time constant, defined by its internal conductivity and heat capacity. In the example of the \emph{left panel} of figure~\ref{fig:model_BB}, this time constant is $\tau_{2}=\SI{20}{s}$. The result, under a \SI{200}{s} period square--wave excitation for the power dissipated in the resistors, is a reduced amplitude, approximately triangle--wave variation of the temperature of the ECCOSORB HR10 emitting surface. The delay between maximum temperature of the copper plate and maximum temperature of the HR10 foam: it is $\sim\SI{8}{s}$, consistent with the delay between maximum temperature of the copper plate and maximum amplitude of the detected signal measured and shown in the \emph{right panel} of figure~\ref{fig:model_BB}. In these conditions, the model gives a ratio between the amplitude of the temperature variation of the copper plate (measured by the thermometer) and the amplitude of the temperature variation of the ECCOSORB HR10 foam: it is $0.83$, which has to be taken into account to estimate the responsivity of the detectors.

Knowing the peak--to--peak amplitude of the temperature of the emitting surface $\Delta T_{F}$, using the Planck formula, we can estimate the peak--to--peak amplitude of the emitted brightness. Comparing to the peak--to--peak amplitude of the detected signal we can finally estimate the optical responsivity of the detectors.

\begin{figure}[!h]
\centering
\includegraphics[scale=0.45]{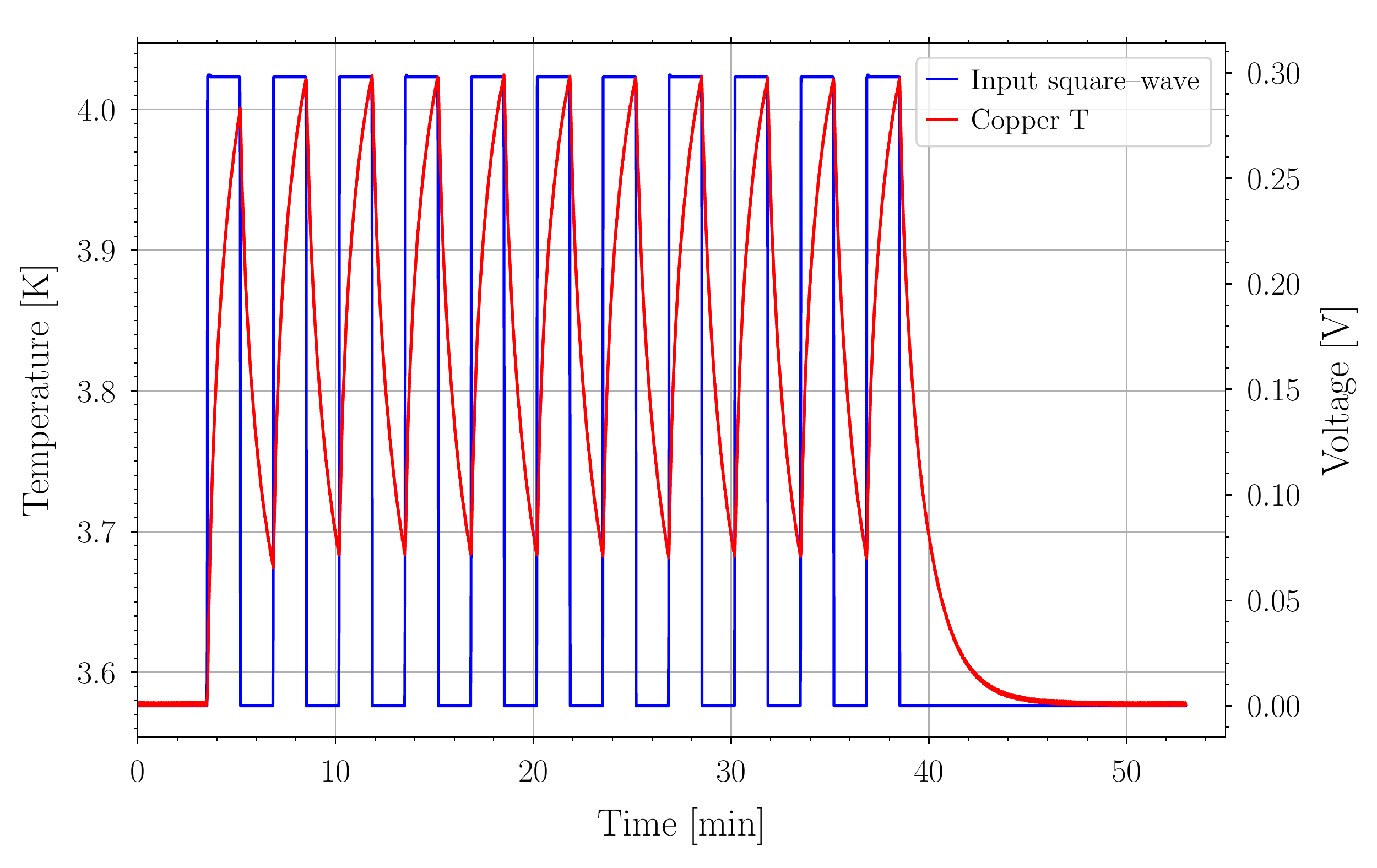}
\caption{\small Measurements of the modulation of the voltage across the heaters (\emph{blue}) and resulting temperature change of the copper flat (\emph{red}), both \emph{versus} time. The input signal is a \SI{135}{\micro W} power pulse, with a period of \SI{200}{s}, resulting in a $\sim\SI{0.33}{K}$ change of the copper temperature.}
\label{fig:meas_BB}      
\end{figure}

The performance of the black--body source has been tested by dissipating intermitting (square--wave) power in the heaters, and monitoring the copper temperature, to select a suitable modulation period. Figure~\ref{fig:meas_BB} shows the best choice of the modulation of the voltage across the heaters and the resulting temperature change of the copper flat: a \SI{5}{mHz} modulation, with an amplitude of the voltage across the heaters of \SI{0.3}{V}, \emph{i. e.} a dissipated power, in the semi--period when the voltage is applied, of $\sim\SI{135}{\micro W}$. The time constant $\tau_1$ of the copper/fiberglass system has been measured from a number of test power pulses with different amplitudes. We found values ranging from \SI{65}{s} to \SI{75}{s}, depending on the amplitude of the power pulses.

\subsection{Measurements}
\label{sec:meas_opt}

\begin{figure}[!h]
\centering
\includegraphics[scale=0.45]{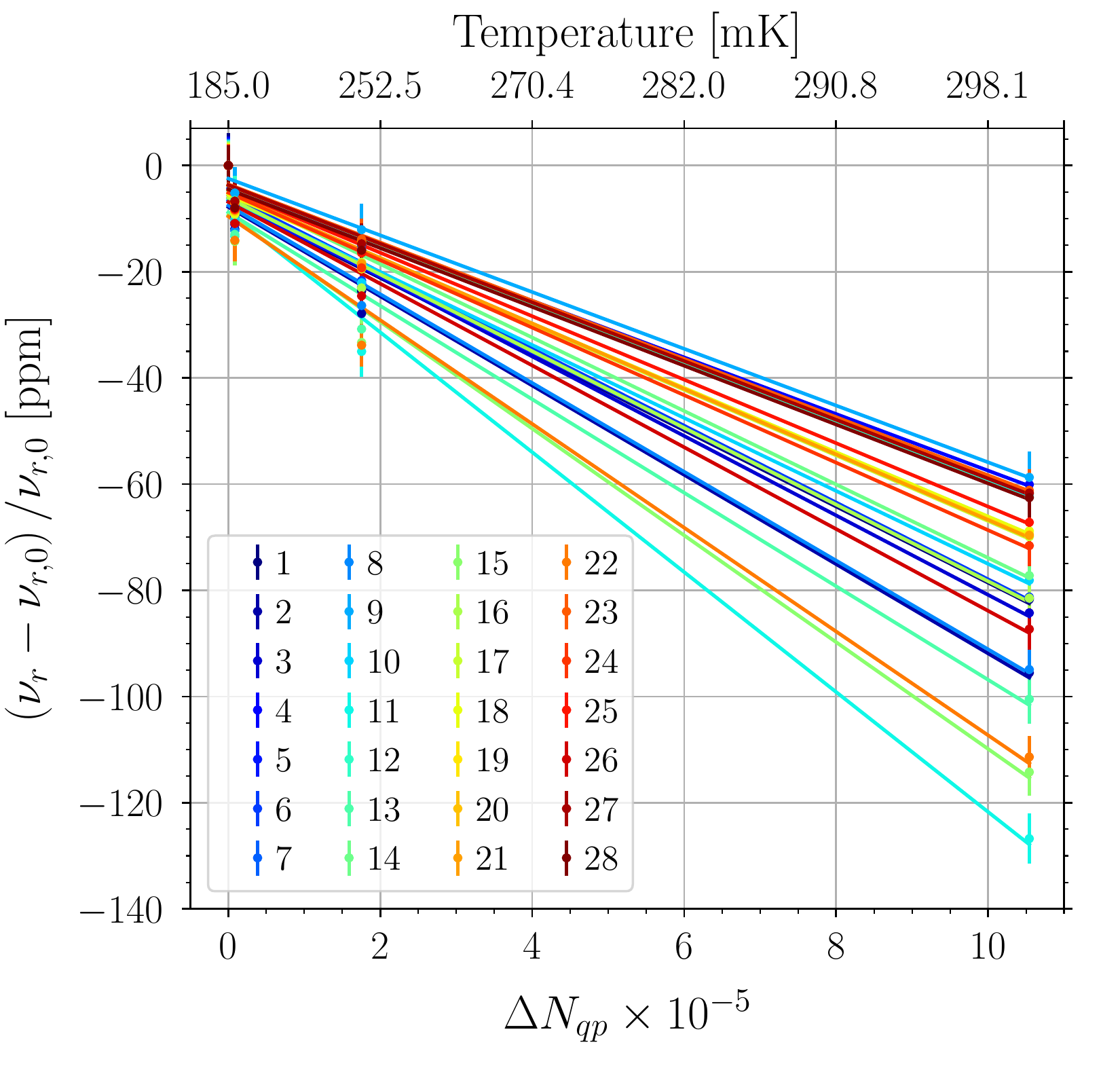}
\caption{\small Trends of the normalised resonant frequency shift with the number of quasi--particle. The \emph{dots with error bars} are the measured data and the \emph{lines} are the linear fits. Different colours are for different pixels.}
\label{fig:elec_resp_tot}      
\end{figure}

The first measurement we have performed was the electrical responsivity under the \SI{3.6}{K} load. Table~\ref{tab:3} collects the array--average values of the electrical parameters measured at \SI{185}{mK}. As described in section~\ref{sec:3}, we have performed a temperature sweep between \SI{185}{mK} and \SI{300}{mK}, see figure~\ref{fig:elec_resp_tot}. We obtained an array--average electrical responsivity in phase

\begin{equation}
\mathcal{R}_{elec,\varphi}= \left(1.16\pm0.30\right)\times10^{13}~\SI{}{W^{-1}}\;,
 \label{eq:3}
\end{equation}
which is compatible with the one measured in the dark conditions in equation~\eqref{eq:2}.
 
 \begin{table}[!h]
 \centering
 \caption{\small Array--average values of the electrical parameters.}
 \vspace{2mm}
\fontsize{10pt}{15pt}\selectfont{
\begin{tabular}{cc}
\hline
\hline
Parameter & Value  \\
\hline
$Q_{c}$ & $\SI{63000}{}\pm\SI{4000}{}$\\
$Q_{tot}$ & $\SI{24000}{}\pm\SI{1000}{}$ \\
$Q_{i}$ & $\SI{38000}{}\pm\SI{2000}{}$\\
resonator dip depth $\left[\SI{}{dB}\right]$& $4.0\pm0.5$\\
\hline
\hline
\end{tabular}}
\label{tab:3} 
\end{table}

\begin{figure}[!h]
\centering
\includegraphics[scale=0.45]{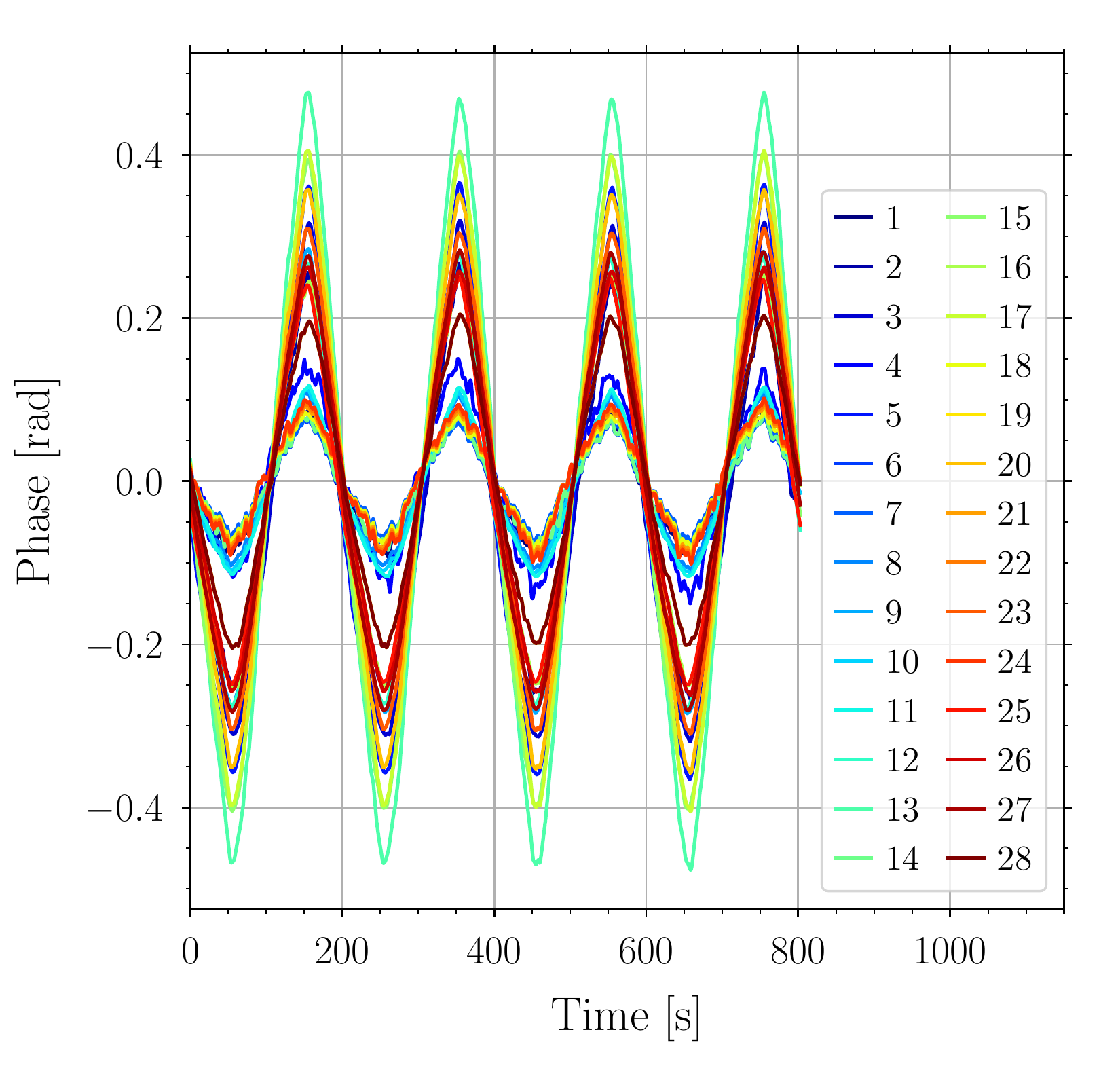}
\includegraphics[scale=0.45]{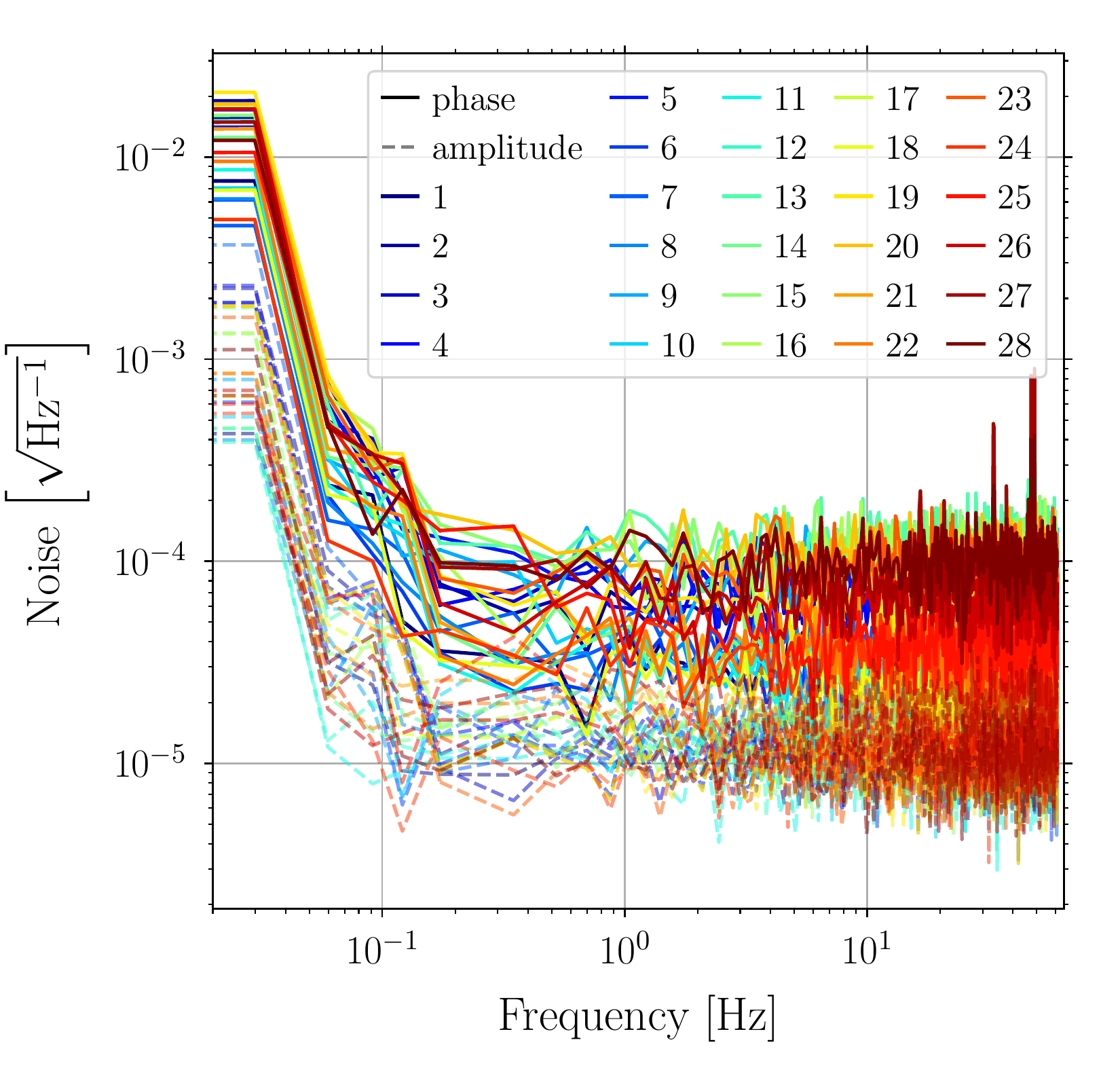}
\caption{\small \emph{Left panel}: Phase time--streams of the detectors when modulating the temperature of the black--body source. Different colours are for different pixels. \emph{Right panel}: Noise power spectra in phase (\emph{solid lines}), compared to noise power spectra in amplitude (\emph{dashed transparent lines}), for all the detectors under a stable \SI{3.6}{K} background. Different colours are for different pixels.}
\label{fig:picco_noise}      
\end{figure}

At this point, in order to evaluate the optical responsivity and NEP, we used the black--body calibrator. The \emph{left panel} of figure~\ref{fig:picco_noise} shows the phase response of all the detectors when modulating the temperature of the black--body source, at an operating temperature of \SI{185}{mK}. As we can see, the signal has the shape of the one predicted in figure~\ref{fig:model_BB}.

The optical responsivity can be estimated by considering that in these measurements the temperature change of the copper flat is $\left(0.33\pm0.01\right)\SI{}{K}$ peak--to--peak and, from the phase shift between the optical signal and the voltage excitation of the heaters, we estimate an amplitude of the emitter surface temperature of about $\left(0.27\pm0.02\right)\SI{}{K}$ peak--to--peak. For a \SI{3.9}{K} average temperature, such a temperature fluctuation produces a power of $\left(8.2\pm0.3\right)\times 10^{-14}~\SI{}{W}$ in a $\lambda^{2}$ throughput at \SI{150}{GHz} with a bandwidth of \SI{23}{GHz} (bandwidth defined by the \SI{150}{GHz}--centered band-pass filter shown in the \emph{right panel} of figure~\ref{fig:HFSS_simu}), and with an in--band emission efficiency of the emitting surface of 0.9.

\begin{figure}[!h]
\centering
\includegraphics[scale=0.5]{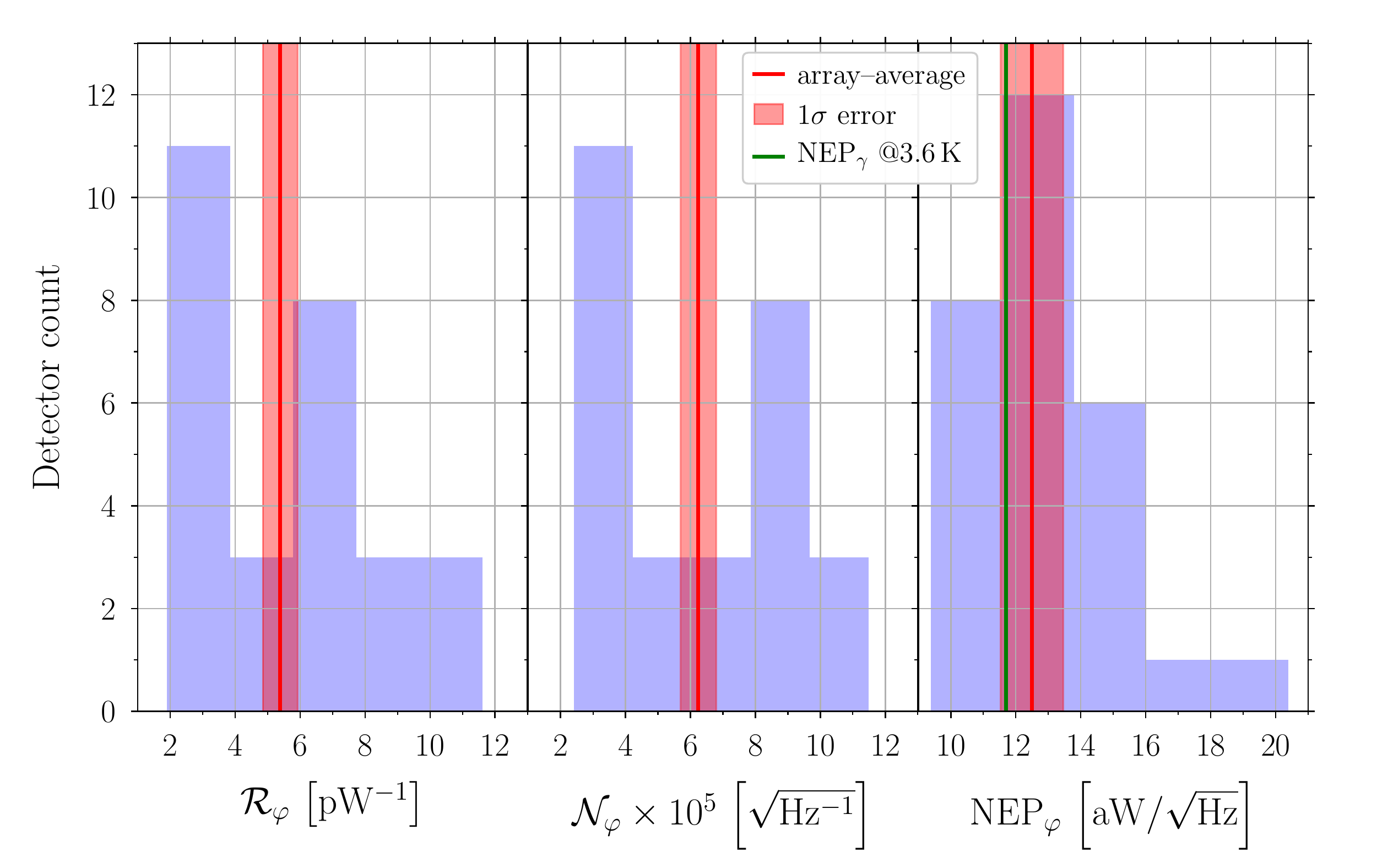}
\caption{\small Histograms of the detector count \emph{versus} the phase optical responsivity (\emph{left panel}), the phase noise (\emph{central panel}) and the phase optical NEP (\emph{right panel}). The \emph{red lines} are the array--average values, the \emph{light red boxes} are the $1\sigma$ errors of the array--average values, and the \emph{green line} is the expected photon noise in the \SI{3.6}{K} black--body background of the measurements.}
\label{fig:histo}      
\end{figure}

The \emph{left panel} of figure~\ref{fig:histo} shows the histogram of the detector count \emph{versus} the optical phase responsivity, where we also report the array--average value and its $1\sigma$ error:

\begin{equation}
\mathcal{R}_{opt,\varphi}=\left(5.39\pm0.55\right)\times 10^{12}~\SI{}{W^{-1}}\;.
 \label{eq:4}
\end{equation}
Here we have assumed that all the detectors are illuminated in the same way by the black--body source, which is reasonable given the extension of the emitting surface, which is larger than the size of the array, and the small distance ($\sim\SI{60}{mm}$) between the emitting surface and the feed--horns apertures.

The \emph{right panel} of figure~\ref{fig:picco_noise} shows the phase noise power spectra for all the detectors, under a stable \SI{3.6}{K} radiative background. The \emph{central panel} of figure~\ref{fig:histo} shows the histogram of the detector count \emph{versus} the phase noise estimated in the white noise regime, precisely for $f>\SI{0.3}{Hz}$. We also report the array--average value of the phase noise and its $1\sigma$ error:

\begin{equation}
\mathcal{N}_{\varphi}=\left(6.04\pm0.62\right)\times 10^{-5}~\SI{}{\sqrt{\rm Hz^{-1}}}\;.
 \label{eq:5}
\end{equation}

The \emph{right panel} of figure~\ref{fig:histo} shows the histogram of the detector count \emph{versus} the phase optical NEP, were we also report the array--average value and its $1\sigma$ error:

\begin{equation}
{\rm NEP}_{opt,\varphi}= \left(1.25\pm0.10\right)\times10^{-17}~\SI{}{W/\sqrt{\rm Hz}}\;.
 \label{eq:6}
\end{equation}
For comparison, the photon noise produced by the radiative background during the measurements (a \SI{3.6}{K} black--body) in this band and in this throughput is NEP$_\gamma \sim 1.17\times10^{-17}~\SI{}{W/\sqrt{\rm Hz}}$. From the histogram, we can say that 20/28 detectors are photon--noise limited and all detectors have a measured NEP lower than twice the photon--noise.

The dispersion of the optical responsivities reflects the dispersion of the electrical responsivities, and is due to two main reasons: the dispersion of the normalised resonant frequency shift with the number of quasi--particles, shown in figure~\ref{fig:elec_resp_tot}, and the dispersion of the values of $Q_{tot}$. Both depend on the \emph{electrical} section of the KID, namely the capacitors and the coupling to the feedline. In fact, all the pixels have the same absorber/inductor design, which means they have the same geometrical and kinetic inductances (which depend on the frequency very weakly). Precisely, the dispersion of the normalised resonant frequency shift with the number of quasi--particle depends, indeed, on the different values of the resonant frequencies, while the dispersion of the values of $Q_{tot}$ depends on the dispersion of the values of $Q_{c}$ (due to a non perfect compatibility between simulation forecasts and measurements) and on the dispersion of the values of $Q_{i}$ due to the different values of the resonant frequencies. 

Comparing the array--average electrical responsivity in equation~\eqref{eq:3}, and the array--average optical responsivity in equation~\eqref{eq:4}, we have obtained an array--average optical efficiency of

\begin{equation}
\varepsilon= \left(46\pm10\right)\%\;.
 \label{eq:7}
\end{equation}
This measured efficiency is the product of the in--band transmission of the \SI{150}{GHz}--centered band--pass filter, $t_{\SI{150}{GHz}}=\SI{70}{\%}$ (\emph{green line} in the \emph{right panel} of figure~\ref{fig:HFSS_simu}), the in--band transmission of the \SI{240}{GHz} low--pass filter, $t_{\SI{240}{GHz}}=\SI{90}{\%}$, and the in--band absorption efficiency of the KIDs, $\eta=\SI{75}{\%}$ (\emph{red line} in the \emph{right panel} of figure~\ref{fig:HFSS_simu}). Therefore, the measured optical efficiency in equation~\eqref{eq:7} is compatible with the expected: $\varepsilon_{\rm expect}=t_{\SI{150}{GHz}}\;t_{\SI{240}{GHz}}\;\eta=\SI{47}{\%}$.

\begin{figure}[!h]
\centering
\includegraphics[scale=0.45]{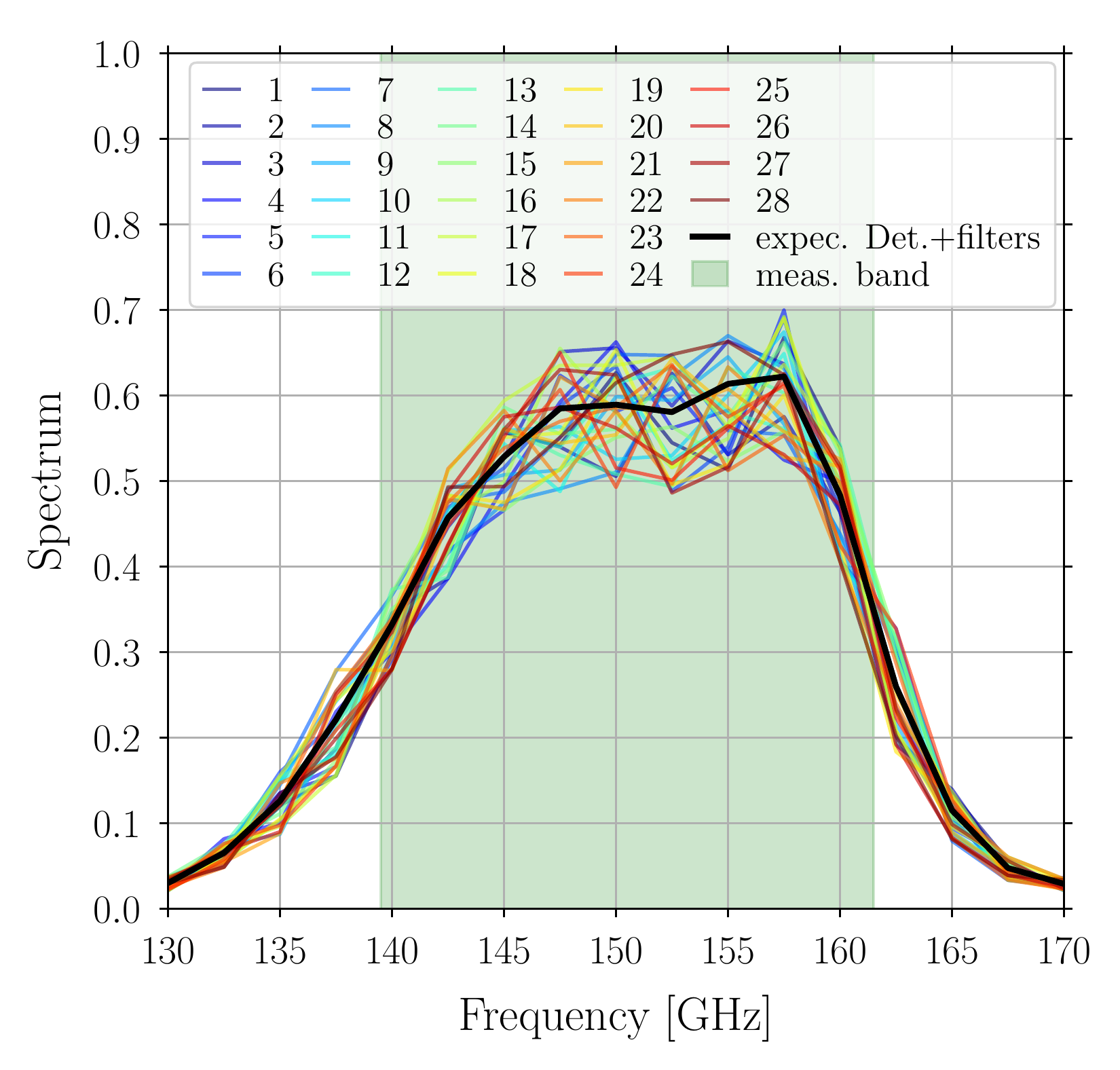}
\caption{\small Measured spectral bands of the detector$+$band--pass filter system. Different colours are for different pixels. The spectra are compared to the expected one (\emph{black}).}
\label{fig:meas_spectra}      
\end{figure}

To confirm the spectral band considered previously, we have measured it by using a 110--\SI{170}{GHz} gunn source and recording the detector response. The gunn has been calibrated by means of a power meter. The source was placed outside the cryostat in such a way that the radiation, coming from the gunn, illuminated the cryostat HDPE (high--density polyethylene) window up to the detector system passing through two thermal blockers and a \SI{300}{GHz} low--pass filter at \SI{40}{K}, a \SI{240}{GHz} low--pass filter at \SI{3.6}{K} and a \SI{150}{GHz}--centered band--pass filter at \SI{1}{K}. The spectrum measurements shown in figure~\ref{fig:meas_spectra} are for the detector$+$band--pass filter system: they are therefore corrected for the transmission of the cryostat window, the thermal blockers, the low--pass filters, and the optical coupling losses.

Finally, we have to mention that this array is an \emph{evolution} of the OLIMPO \SI{150}{GHz} array, which had been operated in a stratospheric environment, but was designed to have lower dynamics and higher responsivity. For the issue of the high--rate of cosmic rays which the detector array would experience in the space environment, we expect a performance similar to the one measured for OLIMPO, and described in \cite{Masi_2019, Paiella_2020}.

\section{One example of space application}
\label{sec:5}

Here we show how a detector array with the performance demonstrated by our prototype is suitable for a future space--mission for precision measurements of spectral distortions of the CMB. 

In a FTS--based measurement, the detector measures simultaneously radiation from the entire measurement band. If the target is the CMB, the spectral coverage of the instrument can start from $\nu_{min}\sim\SI{120}{GHz}$ (the minimum frequency detected by aluminum KIDs) and extend to the maximum frequency of interest for the CMB, $\nu_{max}\sim\SI{600}{GHz}$. 

Since our optical coupling strategy selects the optimal wafer thickness for a given center frequency, we plan to divide the observation band above in sub--bands. This has the additional advantage of avoiding contamination from high--frequency noise at low frequencies. Here we consider the sub--band starting at $\sim\SI{120}{GHz}$ and extending to $\sim\SI{220}{GHz}$, where the absorption of our prototype is still significant (as deducible from the \emph{red line} of figure~\ref{fig:HFSS_simu}). The detector will be single--moded at the low--frequency end of the band (to maximise the angular resolution of the instrument) and multi--moded at the high--frequency end. The throughput will then be $\sim \lambda_{max}^2 = c^2/\nu_{min}^2$ for the entire band covered by the instrument.

\begin{figure}[!h]
\centering
\includegraphics[scale=0.5]{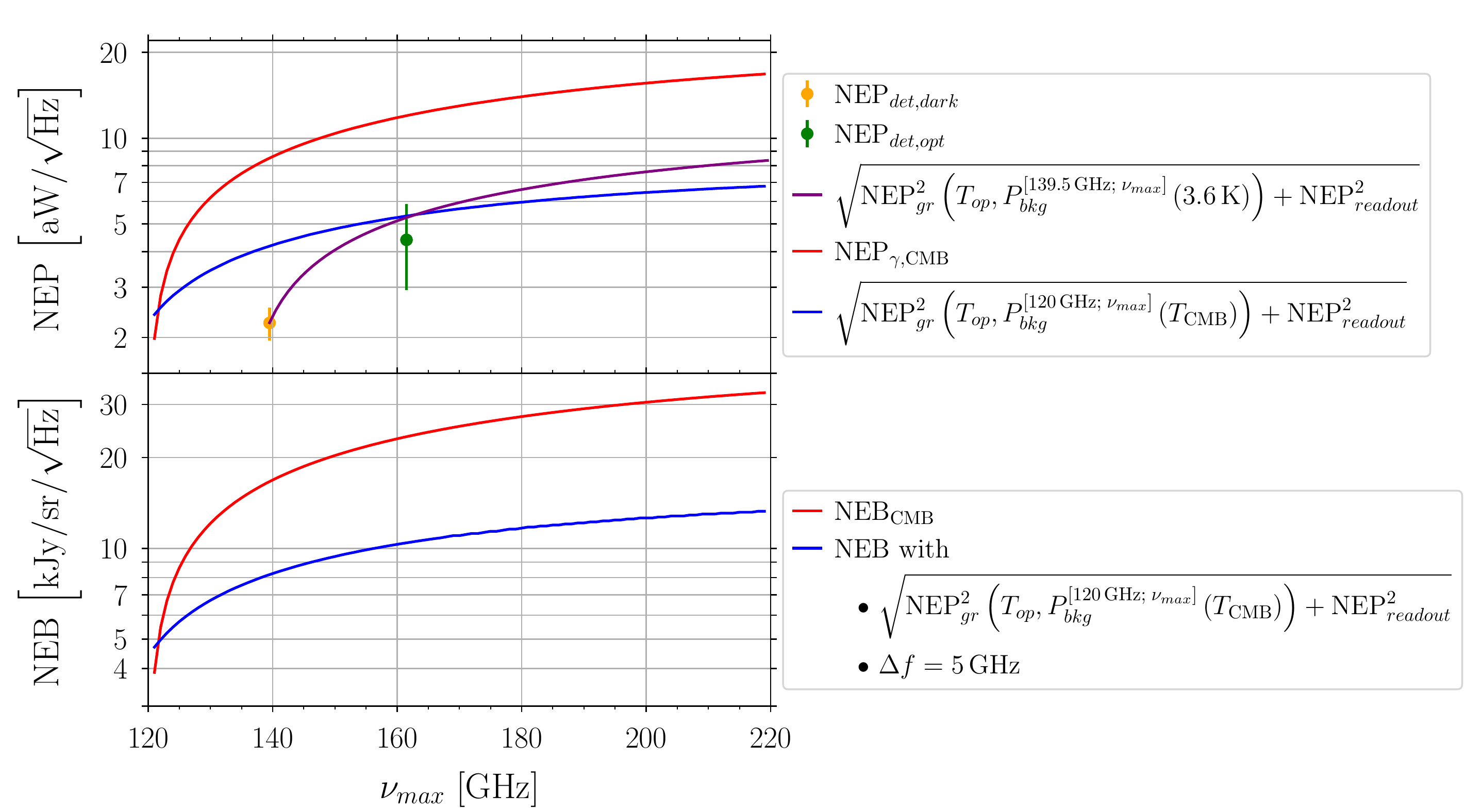}
\caption{\small \emph{Top panel}:  photon noise from the CMB (a \SI{2.725}{K} black--body), for a photometric band starting at $\nu_{min}=\SI{120}{GHz}$, and extending to the maximum frequency reported on the abscissa (\emph{red line}). The \emph{blue line} corresponds to the detector NEP, under the radiative background of the CMB, modelled from the measurements on our detector, as described in the text. The \emph{orange data point} is the measured array--average dark NEP and the \emph{green data point} is the  array-average optical NEP, measured under the radiative background of a \SI{3.6}{K} black--body. The \emph{purple line} represents the detector NEP model for a \SI{3.6}{K} black--body radiative background, in the $\left[\SI{139.5}{GHz};\;\nu_{max}\right]$ spectral band. \emph{Bottom panel}: Noise Equivalent Brightness (NEB) corresponding to the detector NEP model, \emph{blue line}, for a FTS measurement with spectral resolution of \SI{5}{GHz}, with the same spectral coverage. The \emph{red line} is the unavoidable contribution from the photon noise of the CMB, which is dominant.} \label{fig:spacebackground}
\end{figure}

The combination of large throughput and wide frequency coverage implies a significant radiative background from the environment. In a cryogenic space mission aimed at the CMB, with a cold optical system, this is dominated by the \SI{2.725}{K} black--body (plus a small contribution at high frequencies from diffuse interstellar dust in our Galaxy). A suitable detector will be photon--noise limited in this radiative environment. In the \emph{top panel} of figure~\ref{fig:spacebackground}, we compare the photon noise in the band of interest, as a function of $\nu_{max}$, to the intrinsic NEP expected for our detector array, estimated starting from the measurements. In details, we modelled the intrinsic noise of our detectors as:
\begin{equation}
{\rm NEP}_{det} = \sqrt{{\rm NEP}^{2}_{gr}+{\rm NEP}^{2}_{readout}}\;,
\label{eq:NEP_det}
\end{equation}
where ${\rm NEP}_{readout}$ is the noise due to the cold and warm bias/readout electronics and ${\rm NEP}_{gr}$ is the generation--recombination noise of the detector, which depends on the operating temperature, $T_{op}$, and on the background power on the detectors, $P_{bkg}$: 
\begin{equation}
{\rm NEP}_{gr}\left(T_{op},P_{bkg}\right) = 2\Delta\sqrt{\frac{N_{qp}^{th}\left(T_{op}\right)}{\tau_{qp}}+\frac{P_{bkg}}{\Delta}}\;,
\label{eq:NEP_gr}
\end{equation}
with $N_{qp}^{th}\left(T_{op}\right)$ number of thermal quasi--particles at $T_{op}=\SI{185}{mK}$, computed with the BCS theory. In this model, the dark measurements, collected in table~\ref{tab:2}, constrain ${\rm NEP}_{readout}$ by considering ${\rm NEP}_{det,dark}={\rm NEP}_{dark,\varphi}$ and ${\rm NEP}_{gr}={\rm NEP}_{gr}\left(T_{op},0\right)$ (for the dark measurements the contribution of $P_{bkg}$ is indeed negligible), while the optical measurements, described in section~\ref{sec:meas_opt}, serve to verify the model in equation~\ref{eq:NEP_det}, by considering ${\rm NEP}_{det,opt}=\sqrt{{\rm NEP}^{2}_{opt,\varphi}-{\rm NEP}^{2}_{\gamma}\left(\SI{3.6}{K}\right)}$ and
\begin{equation}
{\rm NEP}_{gr}={\rm NEP}_{gr}\left(T_{op},P_{bkg}^{\left[\SI{139.5}{GHz};\;\nu_{max}\right]}\left(\SI{3.6}{K}\right)\right)\;,
\label{eq:NEP_gr_model}
\end{equation}
where $P_{bkg}$ is the power due to a \SI{3.6}{K} black--body in the $\left[\SI{139.5}{GHz};\;\nu_{max}\right]$ spectral band. In the \emph{top panel} of figure~\ref{fig:spacebackground}, the dark measured NEP is plotted at $\nu_{max}=\SI{139.5}{GHz}$ (which means $P_{bkg}=\SI{0}{W}$), while the ${\rm NEP}_{det,opt}$ is plotted at $\nu_{max}=\SI{161}{GHz}$, the upper edge of the optical band during our measurements. Here, we also show the model
\begin{equation}
    \sqrt{{\rm NEP}^{2}_{gr}\left(T_{op},P_{bkg}^{\left[\SI{139.5}{GHz};\;\nu_{max}\right]}\left(\SI{3.6}{K}\right)\right)+{\rm NEP}^{2}_{readout}}\;,
    \label{eq:Mmodel}
\end{equation}
which fits well the experimental points. Therefore, the intrinsic NEP expected for our detector array in deep--space operation, under the sole radiative background of the CMB, is given, according to the model, by  
\begin{equation}
    \sqrt{{\rm NEP}^{2}_{gr}\left(T_{op},P_{bkg}^{\left[\SI{120}{GHz};\;\nu_{max}\right]}\left(T_{\rm CMB}\right)\right)+{\rm NEP}^{2}_{readout}}\;.
    \label{eq:Mmodel_CMB}
\end{equation}
This is also reported in the \emph{top panel} of figure~\ref{fig:spacebackground}.

We also compute the NEB (noise equivalent brightness) for a spectral FTS measurement under the same background, as \cite{2012AA...538A..86D}
\begin{equation}
{\rm NEB} = 0.61 \frac{\sqrt{\displaystyle{\int_{\nu_{min}}^{\nu_{max}} {\rm NEP}^2_{\nu} d\nu}}}{A \Omega \Delta f}
\label{eq:8}
\end{equation}
where $\Delta f$ is the spectral resolution of the measurement and $A \Omega \sim c^2/\nu_{min}^2 $ is the throughput of the detector. This is plotted in  the \emph{bottom panel} of figure~\ref{fig:spacebackground}. From figure~\ref{fig:spacebackground}, it is evident that:
\begin{itemize}
\item For broad band photometric measurements in a deep--space environment, the photon noise of the CMB dominates over the intrinsic noise of our detector, for any reasonable bandwidth starting at 120 GHz.  
\item For FTS-based spectroscopic measurements, the prototype we have developed is suitable for photon--noise limited measurements of spectral distortions of the CMB in a deep--space radiative environment, if a wide frequency band between \SI{120}{GHz} and \SI{220}{GHz} is covered with \SI{5}{GHz} resolution.  
\item In these conditions, a FTS instrument based on this detector array would have the capability of detecting the $y$ distortion of the CMB monopole ($y\sim 2\times 10^{-6}$ \cite{2015PhRvL.115z1301H}, corresponding to a brightness fluctuation of $\SI{-3}{kJy/sr}$ at \SI{140}{GHz}) in a reasonable integration time (signal--to--noise ratio $\sim 20$ in \SI{1}{day}, with a single detector, for each spectral element); lower level distortions can be detected, using the full array, within the few years duration of a space mission of this kind.
\end{itemize}
Additional arrays can be produced with the same technology and similar performance, to extend the spectral coverage up to \SI{600}{GHz}. 

\section{Conclusion}
\label{sec:6}

In the context of the ASI/KIDS project, we have designed, fabricated and characterised, electrically and optically, a 37--pixel array of LEKIDs at \SI{150}{GHz}, completed of its feed--horn $+$ single--mode circular waveguide array, sensitive to the total power of the radiation, optimised to work on--board a satellite mission, cooled at about \SI{185}{mK} and efficiently replicable to populate a large focal plane. 

We have shown the electrical performance of a prototype pixel and the electrical and optical tests of the final array. For the latter, we have developed a black--body calibrator cooled at \SI{3.6}{K}, which allowed us to measure the optical properties in a radiative background representative of a space mission. A large fraction of the pixels of the array are photon--noise--limited in the radiative background of a \SI{3.6}{K} black--body, which is roughly representative of the radiative background of a space mission observing the CMB through a cryogenically cooled optical system. 

\acknowledgments{
\addcontentsline{toc}{section}{Acknowledgements}
We thank the Italian Space Agency for funding the KIDS project with the program \emph{Ideas of new scientific instrumentation for space}. We warmly thank all the colleagues of the \emph{Laboratorio di rivelatori Criogenici} of the Dipartimento di Fisica, \emph{Sapienza} Universit\`a di Roma \& INFN sezione Roma1 for the dilution cryotat used for preliminary tests. 
}

\bibliographystyle{JHEP} 
\addcontentsline{toc}{section}{References}
\bibliography{biblio.bib}

\end{document}